\documentclass[fleqn,usenatbib]{mnras}

\usepackage[normalem]{ulem}
\usepackage{verbatim}

\usepackage{braket}

\usepackage[T1]{fontenc}

\DeclareRobustCommand{\VAN}[3]{#2}
\let\VANthebibliography\thebibliography
\def\thebibliography{\DeclareRobustCommand{\VAN}[3]{##3}\VANthebibliography}

\usepackage{graphicx}	
\usepackage{amsmath}	
\usepackage[dvipsnames]{xcolor}
\usepackage{txfonts}

\newcommand{\rev}[1]{\textcolor{black}{#1}}  
\newcommand{\revmnras}[1]{\textcolor{black}{#1}}

\newcommand{\rc}{r_{\text{c}}}
\newcommand{\tacc}{t_{\text{acc},0}}
\newcommand{\tacct}{t_{\text{acc}}(t)}
\newcommand{\Macc}{\dot{M}_{\rm{acc}}}
\newcommand{\rin}{r_{\rm{in}}}
\newcommand{\alphaDW}{\alpha_{\rm{DW}}}
\newcommand{\alphaSS}{\alpha_{\rm{SS}}}
\newcommand{\tdisp}{t_{\rm{disp}}}
\newcommand{\MaccMd}{\dot{M}_*-M_{D}}
\newcommand{\Sigmac}{\Sigma_{\rm{c}}}

\title{Secular evolution of MHD wind-driven discs: analytical solutions in the expanded $\alpha$-framework}

\author[]{
Benoît Tabone,$^{1}$\thanks{E-mail: tabone@strw.leidenuniv.nl}
Giovanni P. Rosotti,$^{1,2}$
Alexander J. Cridland,$^{3}$
Philip J. Armitage,$^{4,5}$ 
Giuseppe Lodato$^{6}$ 
\\
$^{1}$Leiden Observatory, Leiden University, PO Box 9513, 2300 RA Leiden, The Netherlands \\
$^{2}$School of Physics and Astronomy, University of Leicester, Leicester LE1 7RH, UK \\
$^{3}$Max-Planck-Institut für Extraterrestrische Physik, Giessenbachstrasse 1, D-85748 Garching bei München, Germany\\
$^{4}$Department of Physics and Astronomy, Stony Brook University, Stony Brook, NY 11794, USA\\
$^{5}$Center for Computational Astrophysics, Flatiron Institute, New York, NY 10010, USA \\
$^{6}$Dipartimento di Fisica dell'Università degli Studi di Milano, Via Celoria 16, Milano I-20133, Italy \\
}

\date{Accepted XXX. Received YYY; in original form ZZZ}

\pubyear{2020}

\begin{document}
\label{firstpage}
\pagerange{\pageref{firstpage}--\pageref{lastpage}}
\maketitle

\begin{abstract}
The evolution of protoplanetary discs and the related process of planet formation is regulated by angular momentum transport and mass-loss processes. Over the past decade, the paradigm of viscosity has been challenged and MHD disc winds appear as a compelling scenario to account for disc accretion. In this work, we aim to construct the equivalent of the widely used analytical description of viscous evolution for the MHD wind case. The transport of angular momentum and mass induced by the wind is parameterized by an $\alpha$-like parameter and by the magnetic lever arm parameter $\lambda$. Extensions of the paradigmatic Lynden-Bell and Pringle similarity solutions to the wind case are presented. We show that wind-driven accretion leads to a steeper decrease in the disc mass and accretion rate than in viscous models due to the absence of disc spreading. If the decline of the magnetic field strength is slower than that of the gas surface density, the disc is dispersed after a finite time. The evolution of the disc in the $\MaccMd$ plane is sensitive to the wind and turbulence parameters. A disc population evolving under the action of winds can exhibit a correlation between $\dot{M}_*$ and $M_D$ depending on the initial conditions. The simplified framework proposed in this work opens to a new avenue to test the effectiveness of wind-driven accretion from the observed disc demographics and constitutes an important step to include wind-driven accretion in planet population synthesis models.
\end{abstract}

\begin{keywords}
accretion - accretion discs - MHD disc winds – protoplanetary discs – submillimetre: planetary systems
\end{keywords}



\section{Introduction}

Understanding why proto-planetary discs accrete is a necessary step to build any successful planet formation theory (e.g., \citealt{2016JGRE..121.1962M}). Yet, even after decades of study this issue remains elusive.



Accretion is a common phenomenon and almost all the known proto-planetary discs, as identified by their infra-red excess, show some sign of accretion (see \citealt{2016ARA&A..54..135H}, and references therein). In the conventional, "viscous" view, popularised\footnote{Though the notion that turbulence is connected to angular momentum transport pre-dates these works, see \citet{1981ARA&A..19..137P} for a historical perspective.} by \citet{1973A&A....24..337S} and \citet{1974MNRAS.168..603L}, accretion is ultimately connected with turbulence. At a macroscopic level, turbulence acts as an effective viscosity and redistributes the angular momentum in the disc, transporting it outwards with a small fraction of the mass (a process called ``viscous spreading''). This transport allows the bulk of the mass to move inwards and eventually fall onto the star. Unfortunately, to what extent discs should be turbulent is a long-standing problem (see \citealt{2014prpl.conf..411T} for a review); a mechanism called magneto-rotational instability \citep[MRI,][]{1991ApJ...376..214B} is thought to be the best candidate, but it is currently unclear whether it can generate enough turbulence to explain the observed accretion rates in planet-forming discs. \rev{Alternatively, pure hydrodynamical instabilities such as the gravitational instability, \citep[GI,][]{2016ARA&A..54..271K} or the vertical shear instability, \citep[VSI,][]{2013MNRAS.435.2610N} could also enhance turbulence}.  Usually, following \citet{1973A&A....24..337S}, our ignorance is hidden in the dimensionless parameter $\alpha$ introduced by \citet{1973A&A....24..337S}.

It should be mentioned that, besides its global importance for disc evolution, as a local phenomenon turbulence plays a role in almost any area of planet formation. Just to name a few examples, turbulence affects the efficiency of gas \citep{2013ApJ...770..120B} and dust (e.g., \citealt{2015A&A...582A.112B,2018A&A...615A.178O}) accretion onto forming planets, how discs responds to planets \citep{2012ARA&A..50..211K,2018ApJ...869L..47Z}, the vertical mixing of molecular species \citep{2011ApJS..196...25S}, the importance of fragmentation for dust evolution \citep{2007A&A...466..413O,2012A&A...539A.148B}, and many other processes. 

There is however an alternative to explain accretion, which has received significant attention in the last years, namely the idea that a net vertical magnetic field anchored in the disc could launch a wind that extracts the angular momentum (see e.g. the seminal paper of \citealt{1982MNRAS.199..883B}, and \citealt{2020arXiv200715967L} for a recent review). In this way, the wind exerts a torque on the disc and forces the material in the disc to spiral inward \citep{1997A&A...319..340F}. Because angular momentum is not transported at large radii but removed vertically, there is no viscous spreading. Finally, this scenario does not require any (or only little) turbulence. If correct, this scenario paints a very different picture of discs from the viscous one.

Both redistribution of angular momentum through turbulence (apart from hydrodynamical instabilities) and extraction through winds are ultimately regulated by the magnetic field and its coupling to the gas. Given the cold, low-ionisation conditions of proto-planetary discs, studying in detail the efficiency of these mechanisms requires the use of non-ideal magneto-hydrodynamics (MHD) simulations \citep{2011ARA&A..49..195A}. In fact, numerical simulations taking into non-ideal MHD terms show that in large regions of the disc, called dead-zones, MRI turbulence is suppressed \citep[$\gtrsim 1$~au ;][]{1996ApJ...457..355G,2011ApJ...736..144B}. In this regions, MHD disc winds appear to be a compelling process to drive efficient disc accretion \citep{2013ApJ...769...76B}. However, there are significant numerical challenges in conducting these simulations, connected with stringent requirements in terms of spatial resolution, the time-stepping algorithm, the constraint that the magnetic field divergence should vanish, and the need to couple the simulations with sub-grid micro-physics to estimate the importance of the non-ideal MHD terms \citep[e.g.,][]{2017A&A...600A..75B,2017ApJ...845...75B,2019ApJ...874...90W}. All these factors mean that, while the avenue of MHD simulations is certainly of fundamental importance, it is an avenue that needs to be supported and complemented in other ways. 

Today, thanks to the new generation of telescopes, we now have observational programs that have measured global disc properties for large samples. ALMA has surveyed several star forming regions, providing large samples of disc sub-mm fluxes \citep{2014ApJ...784...82M,2016ApJ...828...46A,2016ApJ...831..125P,2016ApJ...827..142B,2017AJ....153..240A,2017ApJ...851...83C,2018ApJ...860...77E,2019A&A...626A..11C,2019MNRAS.482..698C,2020AJ....160..248A}, a proxy for the disc mass, and for a more limited sub-sample disc radii \citep[][]{2017ApJ...851...85B,2018ApJ...859...21A}. In parallel, surveys at optical wavelengths have measured the properties of the central star and its mass accretion rates \citep{2015A&A...579A..66M,2017A&A...604A.127M,2017A&A...600A..20A,2020A&A...639A..58M}. These data are invaluable to provide constraints to the numerical studies that seek to explain why discs accrete starting from fist principles. Indeed, in the recent years a number of studies have analysed the survey data to provide these constraints \citep{2016A&A...591L...3M,2017MNRAS.468.1631R,2017MNRAS.472.4700L,2017ApJ...847...31M,2020A&A...640A...5T}. These studies have \revmnras{determined} reasonable ranges for the $\alpha$ parameter; however, they have almost exclusively considered only the viscous case.

The reason why almost no study based on disc demographics has provided constraints for the wind scenario is because of the lack of simple analytical solutions, equivalent to those discovered by \citet{1974MNRAS.168..603L}. Although 1D global evolutionary models for MHD winds have been published \citep{2013ApJ...778L..14A,2016ApJ...821...80B,2016A&A...596A..74S,2017ApJ...845...31H,2018MNRAS.475.5059K,2019ApJ...879...98C}, a true equivalent of \citet{1974MNRAS.168..603L} is still missing. While clearly idealised, these solutions are simple and powerful: they permit to quantify the average efficiency of angular momentum transport and are a very useful tool to enable large population studies. These solutions do not make any physical assumption regarding the exact nature of the mechanism at the origin of angular momentum transport; they do not tell us \textit{why} the disc is viscous, but they nevertheless tell us what happens \textit{if} the disc is viscous. The purpose of this paper is to provide a parameterization of the efficiency of angular momentum removal by MHD disc winds akin to \citet{1973A&A....24..337S} and, with this in hand, to provide simple, analytical, self-similar solutions to the disc evolution equations akin to \citet{1974MNRAS.168..603L}. We will see in particular how the wind model requires two, rather than only one, free parameters. One of the two parameters is the equivalent of $\alpha$ and permits to identify immediately if viscosity or winds dominate the transfer of angular momentum. We will then discuss the comparison with the available data from disc surveys that could be conducted to constrain these parameters. We leave more detailed comparisons to future papers. 

This paper is structured as follows. In Sec. \ref{sec:model} the basic equations and the adopted parameterization of the wind torque and the wind mass-loss rate is summarized. We then present the steady-state solution and our similarity solutions in Sec. \ref{sec:anal-sol}, including the main features of the solutions in the $\MaccMd$ plane. The main observational diagnostics that can be used to constrain the wind and turbulence related parameters are then briefly reviewed, and our model is discussed in light of the recent non-ideal MHD simulations and existing disc evolution models in Sec. \ref{sec:disscu}. Our findings are summarized in Sec. \ref{sec:conclusions}.

%
%
%
%
%

\section{Model}
\label{sec:model}

\subsection{Basics}

\begin{figure}
	\includegraphics[width=\columnwidth]{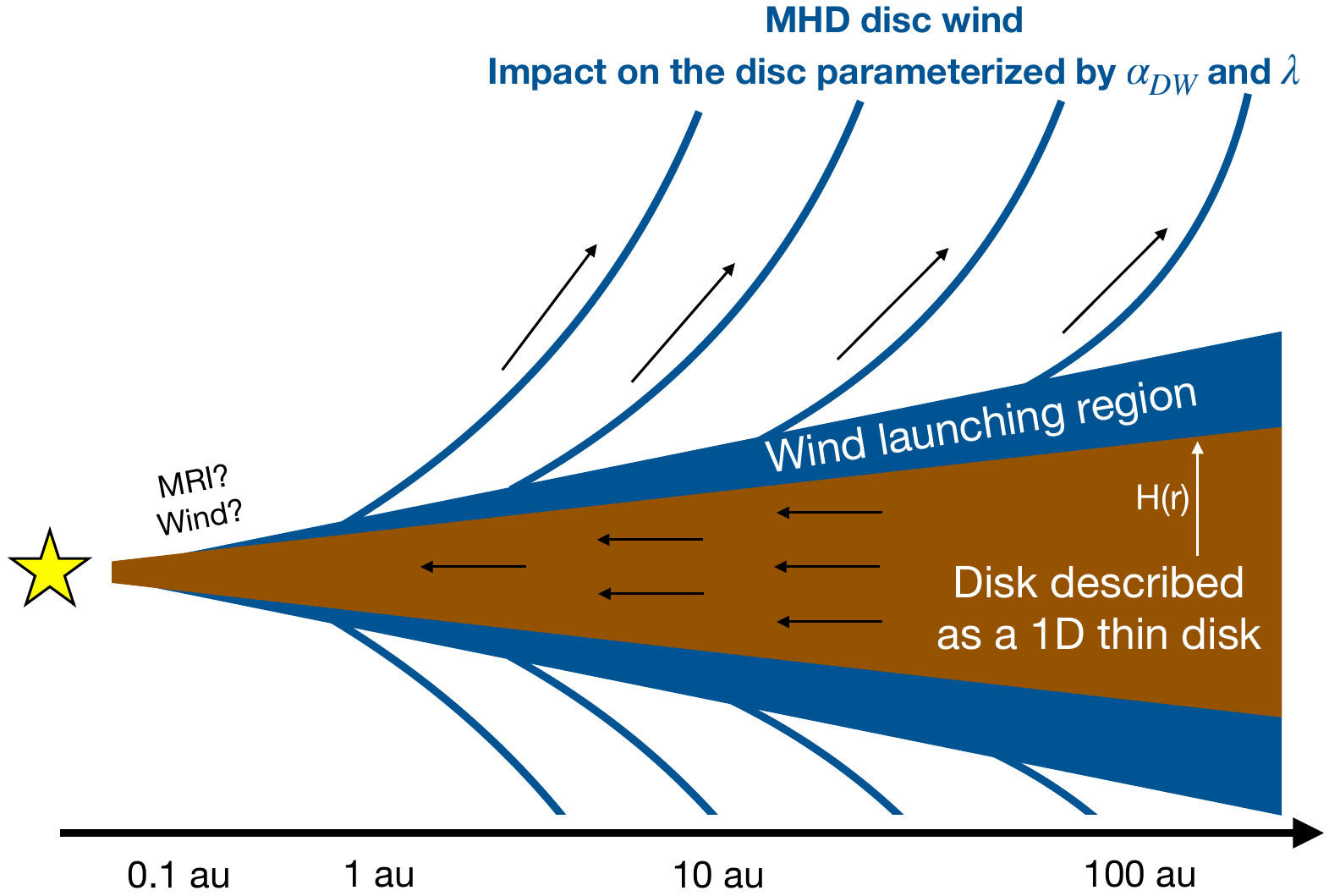}
    \caption{Schematic view of the disc evolution model explored in this work. Disc accretion is driven by an MHD disc wind which extract mass and angular momentum from the disc and by turbulence. We model the disc as a 1D thin disc and describe the impact of the MHD disc wind on the secular evolution of the disc using a Sharura-Sunyaev like parameter denoted as $\alphaDW$ and the magnetic lever arm parameter of the wind denoted as $\lambda$. This simple parameterization allows us to construct analytical solutions for the secular evolution of the disc. 
    }
    \label{fig:schematics}
\end{figure}

The disc is assumed to be geometrically thin and vertically isothermal with $c_s$, the sound speed (see schematic view in Fig. \ref{fig:schematics}). An MHD disc wind transporting mass and angular momentum is launched from a typical scale height of $H_W \sim 2 H$, where $H \equiv c_s/\Omega$ is the hydrostatic scale height of the disc. In this subsection, we also consider a possible radial transport of angular momentum due to e.g., MRI turbulence. The disc is treated in a 1D approach by vertically averaging disc quantities over $\mid z \mid< H_W$. The evolution of the surface density profile is the result of the angular momentum transported radially by turbulence, the angular momentum extracted vertically by the MHD disc wind, and the mass-loss induced by the latter (see Appendix \ref{app:eq-basics} ):

\begin{equation}
\begin{split}
       \frac{\partial \Sigma}{\partial t} = \frac{2}{r}\frac{\partial}{\partial r} \left\{ \frac{1}{r\Omega} \frac{\partial}{\partial r} \left(r^2  \int_{-H_W}^{+H_W} T_{r\phi} dz \right)\right\}
       + \frac{2}{r} \frac{\partial}{\partial r}\left\{\frac{r | T_{z\phi}|^{+H_W}_{-H_W}}{\Omega} \right\}
       - \dot{\Sigma}_W ,
\end{split}
\label{eq:master-eq}
\end{equation}
where
\begin{equation}
\begin{split}
       T_{r\phi} \equiv \braket{\rho \varv_r \delta \varv_{\phi}  - B_r B_{\phi}/4\pi}
\end{split}
\label{eq:Trphi}
\end{equation}
is the time-averaged radial stress tensor describing the radial transport of angular momentum,
\begin{equation}
\begin{split}
       T_{z\phi} \equiv \braket{\rho \varv_z \delta \varv_{\phi} - B_z B_{\phi}/4\pi }   
\end{split}
\label{eq:Tzphi}
\end{equation}
is the time-averaged vertical stress tensor describing the extraction of angular momentum by the MHD disc wind, and $\dot{\Sigma}_W$ is the mass-loss rate per unit surface induced by the wind. The conventional notations for a cylindrical coordinate system is used here. $\Omega = \sqrt{GM_*/r^3}$ is the Keplerian orbital frequency around the young star of a mass $M_*$, and $\delta \varv_{\phi} = \varv_{\phi} - r \Omega$ is the deviation of the rotation velocity to the Keplerian velocity.

The physical quantities $T_{r\phi}$, $T_{z\phi}$, and $\dot{\Sigma}_W$ describe the effective impact of the transport of angular momentum and mass-loss on the disc surface density. The value of these quantities ultimately depends on the physical and chemical structure of the disc (strength and topology of the magnetic field, temperature, density, grain charge and sizes, ionization...). Considering the large uncertainties plaguing models that treat the dynamics of the disc and the wind consistently, we propose here to follow the approach of \citet{1973A&A....24..337S} and parameterize the aforementioned terms using a minimum number of parameters and theoretical preconceptions. 

\subsection{Turbulent and wind torque}
Using our notations, the Shakura-Sunyaev $\alphaSS$-parameter is defined as\footnote{Some authors also adopt a definition of $\alphaSS$ without the $2/3$ prefactor \citep[e.g.][]{2016A&A...596A..74S}} 
\begin{equation}
     \alphaSS \equiv \frac{2}{3} \frac{\int T_{r\phi} dz}{\Sigma c_s^2} \rev{= \frac{2}{3 \sqrt{2 \pi}} \frac{\int T_{r\phi} dz}{H P_0} ,} 
     \label{eq:alpha-SS}
\end{equation}
where $P_0$ is the mid-plane thermal pressure \footnote{With our definition of the hydrodynamical scale height as $H\equiv c_s/\Omega$, the mid-plane thermal pressure is $P_0 = c_s\frac{\Omega \Sigma}{\sqrt{2\pi}}$, and the mid-plane density is $\rho_0 = \frac{\Omega \Sigma}{\sqrt{2\pi} c_s} $.}.
This definition leads to a local accretion rate driven by turbulence of \citep{1981ARA&A..19..137P}
\begin{equation}
    \Macc^{visc}(r) = \frac{6 \pi}{r\Omega}\frac{\partial}{\partial r}(\Sigma c_s^2 \alphaSS r^2).
    \label{eq:Maccvisc}
\end{equation}

By analogy, we normalize the wind torque by the mid-plane thermal pressure and define the dimensionless $\alphaDW$ parameter as
\begin{equation}
    \alphaDW \equiv \frac{4}{3} \frac{r | T_{z\phi}|^{+H_W}_{-H_W}}{\Sigma c_s^2} =  \frac{4}{3\sqrt{2 \pi}} \frac{| T_{z\phi}|^{+H_W}_{-H_W}}{\epsilon P_0},
    \label{eq:alpha-DW}
\end{equation}
where $\epsilon \equiv H/r$ is the disc aspect ratio. With this definition, the local accretion rate driven by the MHD disc wind is
\begin{equation}
    \Macc^{DW}(r) = \frac{ \textcolor{black}{3} \pi \Sigma c_s^2 \alphaDW}{\Omega}.
    \label{eq:MaccW}
\end{equation}
In the presence of turbulence and MHD disc wind, the fraction of mass-accretion driven by the wind compared to that driven by viscosity is roughly the ratio of the $\alpha$-parameters
$\Macc^{DW}/\Macc^{visc} \simeq \alphaDW/\alphaSS$. Therefore, the comparison between the paradigmatic viscous disc model and the wind-driven disc model presented in this paper is greatly facilitated, without loss of generality. This constitutes the main advantage of our definition compared with other conventions adopted in the literature (see Sec. \ref{subsubsec:previous-models} and Appendix \ref{app:alphaDW-def}). We shall already note that numerical simulations and analytical models show that $\alphaDW$ roughly scales with the magnetization of the disc, commonly quantified by the mid-plane $\beta_0$ parameter (see Sec. \ref{subsec:numerical-simu}).

\subsection{Local mass loss rate}
In order to re-write the local mass-loss rate $\dot{\Sigma}_{W}$ in terms of $\alphaDW$ and disc quantities in Eq. (\ref{eq:master-eq}), we use the magnetic lever arm parameter defined as
\begin{equation}
\lambda \equiv \frac{L}{r \Omega(r)},
\end{equation}
where $L$ is the total specific angular momentum carried away along the MHD disc wind streamline anchored at $r$ (in the form of both matter rotation and magnetic torsion). This widely used parameter, first introduced by \citet{1982MNRAS.199..883B} quantifies the ratio of extracted to initial specific angular momentum and can be observationally constrained (see Sec. \ref{subsec:obs-lambda}). For a MHD disc winds, powered by accretion it can be shown that $\lambda>3/2$. The conservation of angular momentum \rev{(see Eq. (\ref{eq:sigmaW-cons-AM}))} gives:
\begin{equation}
\dot{\Sigma}_{W} = \frac{ 3 \alphaDW c_s^2}{\textcolor{black}{4} (\lambda-1) \Omega r^2} \Sigma = \frac{1}{2(\lambda-1)}\frac{ \Macc^{DW}}{\revmnras{2} \pi r^2}, 
\label{eq:mass-loss}
\end{equation}
This approach is similar to that adopted by \citet{2020A&A...633A...4K}.

Equation (\ref{eq:mass-loss}) shows that $\lambda$ can be considered as the efficiency of the wind to drive accretion: the higher $\lambda$, the fewer mass is required to be launched to sustain an accretion rate of $\Macc^{DW}$.

\subsection{Master equation}
The impact of the MHD disc wind on the disc evolution is thus controlled by two local parameters: $\alphaDW$, which quantifies the angular momentum extracted by the wind, and $\lambda$, the magnetic lever arm parameter, which ultimately determines the mass-loss rate of the wind. Injecting Eqs. (\ref{eq:alpha-SS}), (\ref{eq:alpha-DW}), and (\ref{eq:mass-loss}) into Eq. (\ref{eq:master-eq}) gives the master equation:

\begin{equation}
\begin{split}
       \frac{\partial \Sigma}{\partial t} = \frac{3}{r}\frac{\partial}{\partial r} \left\{ \frac{1}{r\Omega} \frac{\partial}{\partial r} \left(r^2\alphaSS \Sigma c_s^2 \right) \right\}
       + \frac{3}{\textcolor{black}{2} r} \frac{\partial}{\partial r}\left\{ \frac{\alphaDW \Sigma c_s^2}{\Omega} \right\}
       - \frac{3 \alphaDW \Sigma c_s^2}{ \textcolor{black}{4} (\lambda-1) r^2 \Omega} .
\end{split}
\label{eq:master-eq-final}
\end{equation}
This equation is a generalization of the standard disc equation. 
Comparing the first two RHS terms, one can already identify the main difference between turbulence-driven accretion and wind-driven accretion: the second term, which describes the wind-driven accretion is a first order differential term in space, corresponding to an advection term in Eq (\ref{eq:master-eq-final}). This means that the accretion velocity due to the wind
\begin{equation}
\varv_{DW} = \frac{3}{\textcolor{black}{2}} \alphaDW \epsilon c_s
\label{eq:acc-velo-DW}
\end{equation}
does not depend on any shear, in contrast to viscous accretion (first RHS term). \rev{We note that $\varv_{DW}$ should be considered as an average over the disc scale height. Numerical simulations show that the accretion can proceed in a \revmnras{thin} layer of gas in the disc atmosphere \citep[e.g.,][]{2013ApJ...769...76B}. In that case, the advection speed in the upper layers can be near sonic velocity ($\simeq c_s$) but the average accretion velocity $\varv_{DW}$ would still be much smaller.}  


\subsection{Assumptions on $\alphaSS$, $\alphaDW$, and $\lambda$}
The master equation Eq. (\ref{eq:master-eq-final}) does not assume anything about the MHD disc wind or about the level of turbulence \rev{as it is based on a parameterization of their local impact on the disc}. As such, Eq. (\ref{eq:master-eq-final}) can be solved numerically, using any prescription for the spatial and temporal evolution of $\alphaSS$, $\alphaDW$, and $\lambda$. 

In this work, we adopt simplifying prescriptions for these parameters, that allows us to find analytical solutions of the evolution of the disc. In particular, we assume that $\alphaSS$, $\alphaDW$, and $\lambda$ are constant in space. Further assumptions about the time evolution of $\alphaDW$ are made for the self-similar solutions (see Sec. \ref{subsec:classes-solution}). The temperature of the disc follows a power-law with $T(r)\propto r^{-1/2}$ \citep{1987ApJ...323..714K,1997ApJ...490..368C} and is constant in time. The solutions given below are further extended in Appendix \ref{app:gamma-solutions} to include $\alpha$-parameters and temperature profiles that vary as a power-law with radius with arbitrary power-law indexes. 

In order to ensure an explicit transition in the analytical expressions from pure wind accretion and pure turbulent accretion, we define 
\begin{equation}
\tilde{\alpha} \equiv \alphaDW+\alphaSS,
\label{eq:alpha-def}
\end{equation}
the $\alpha$-parameter that quantifies the total torque exerted by the MHD disc wind and turbulence, and
\begin{equation}
\psi \equiv \frac{\alphaDW}{\alphaSS},
\label{eq:psi-def}
\end{equation}
a parameter that quantifies the relative strength between the radial and the vertical torque \footnote{We note that $\psi$ is not exactly the ratio of the vertical to the radial torque, that is usually denoted as $\Lambda$ \citep[e.g.,][]{1993A&A...276..637F}.}. 



\section{Analytical solutions}
\label{sec:anal-sol}

In this section, we generalize the analytical solutions proposed by \citet{1974MNRAS.168..603L} to MHD wind-driven discs using two prescriptions for the time evolution of $\alphaDW$. As a preliminary, the study of the steady-state solutions allows us to define two additional key physical quantities, namely the local mass ejection index $\xi$ and the global mass ejection-to-accretion ratio $f_M$.

\subsection{Steady-state}
\label{subsec:steady-solution}



We first assume steady-state $(\partial_t = 0)$ in Eq. (\ref{eq:master-eq-final}). By construction, $\alphaDW$, $\alphaSS$, and $\lambda$ are constant in time. 

\subsubsection{Surface density profile}

\begin{figure}
	\includegraphics[width=\columnwidth]{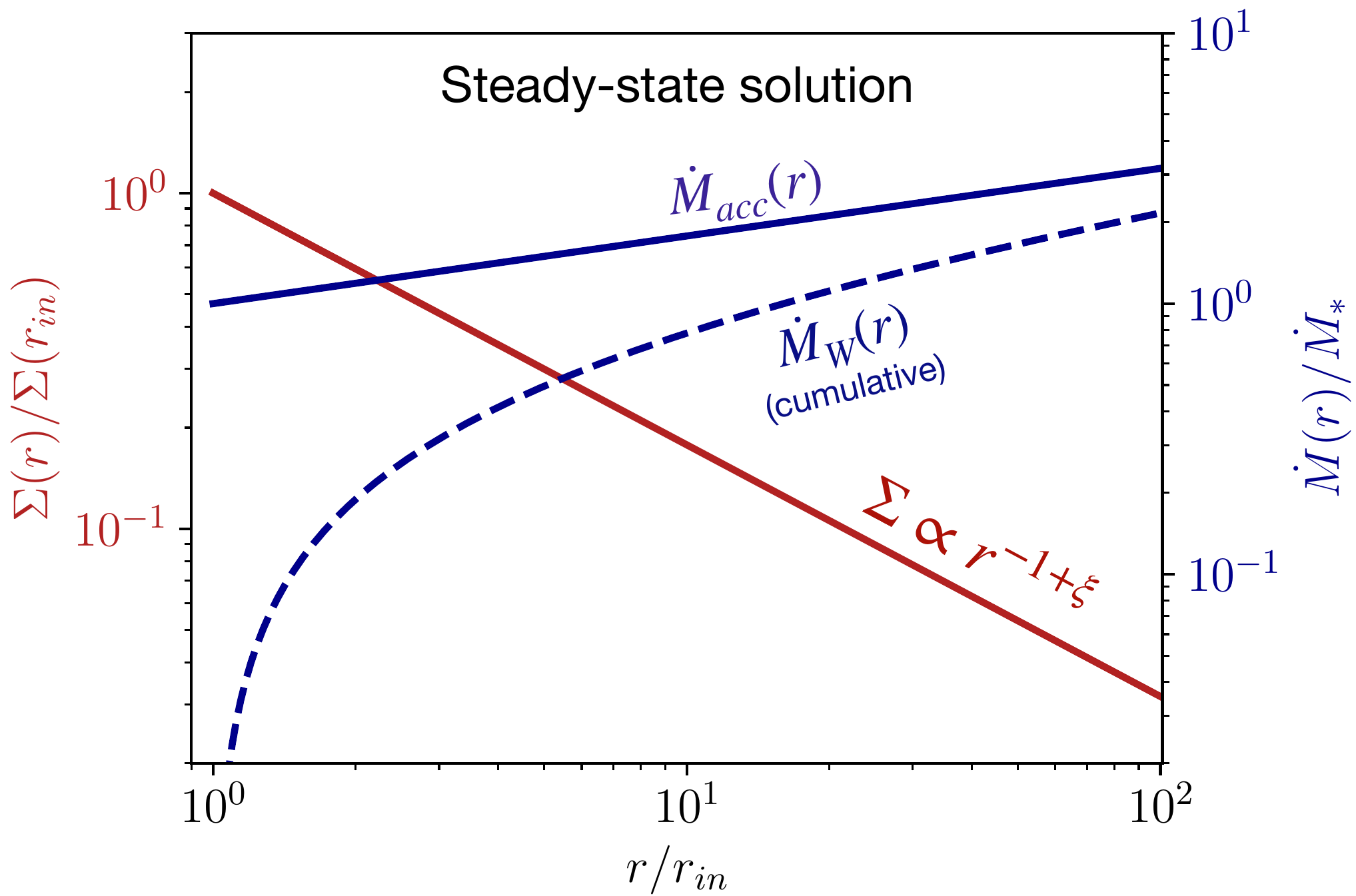}
    \caption{Steady-state solution. Radial profile of the surface density (Eq. (\ref{eq:steady-state})), of the local accretion rate $\dot{M}_{acc}(r)$ (Eq. (\ref{eq:Macc-local})), and of the cumulative mass loss rate $\dot{M}_{W}(r)$ integrated from the inner radius $r_{in}$ to $r$ (Eq. (\ref{eq:MW-local})). The profiles are set by the value of $\xi$ which depends on $\lambda$ and $\psi$ as shown in Fig. \ref{fig:xi} and given in Eq. (\ref{eq:xi}). In this example, we set $\lambda=3$ and $\psi=\infty$ (pure wind case), corresponding to $\xi=0.25$. The mass loss rate and the accretion rate are normalized to the accretion rate at the inner radius. The accretion rate decreases toward $r_{in}$ as mass is lost along with the accretion.
    }
    \label{fig:steady-state-solution} 
\end{figure}

\begin{figure}
	\includegraphics[width=\columnwidth]{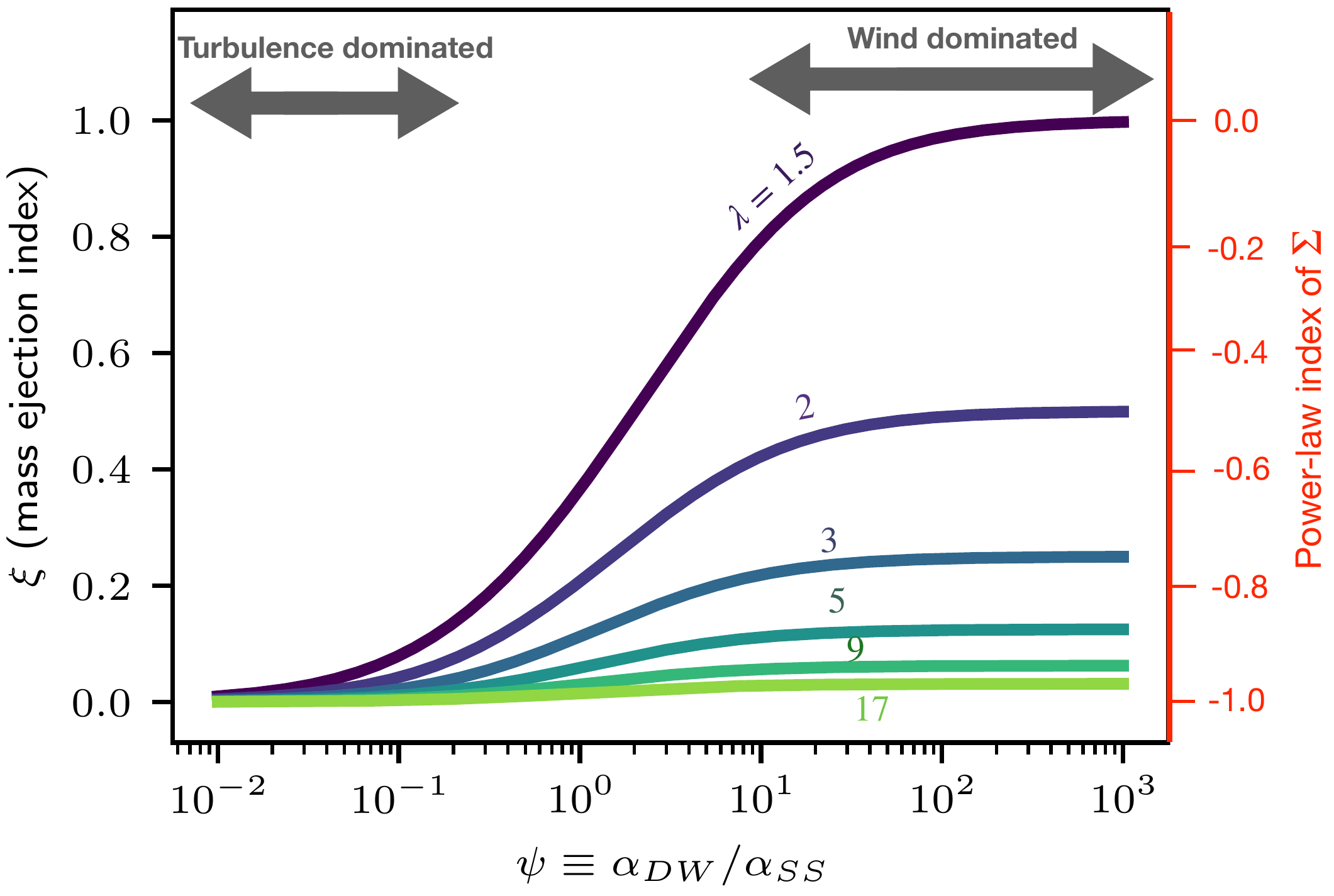}
    \caption{Mass ejection index $\xi$ as a function of $\psi$ for various values of $\lambda$ indicated along each curve (see Eq. (\ref{eq:xi})). $\xi$ controls the slope of $\Sigma$ in the steady-state solutions (see Fig. \ref{fig:steady-state-solution}) and the power-law index of the inner part of the self-similar solution (see Fig. \ref{fig:schematic-similarity}). $\xi$ also set the total wind mass-loss rate provided the radial extent of the wind launching region $r_c/r_{in}$ (see Eqs. (\ref{eq:xi-def-steady-state}) and  (\ref{eq:fM-def})).}
    \label{fig:xi} 
\end{figure}

In the pure viscous case ($\psi = 0$), the surface density follows a simple power-law profile
\begin{equation}
    \Sigma(r) \propto r^{-1}.
    \label{eq:steady-turb}
\end{equation}
The power-law index of -1 is the result a constant accretion velocity of $\varv_{\nu} = \frac{3}{2} \alphaSS \epsilon c_s$: as a parcel of accreting gas moves inward at constant speed, it contracts following a geometrical factor proportional to $r^{-1}$. \rev{This property follows from the choice of the radial profile of $\alphaSS$ and $T$.}

For non-vanishing $\psi$, the total accretion velocity is also constant across the disc. 
One could then expect to recover a surface density profile with a power-law index of -1 as in the viscous case. 
However, wind accretion is also accompanied by a mass-loss that decreases the mass of the advected parcel of gas, resulting in flattening the radial profile of $\Sigma$.
Specifically, adopting a power-law profile (see Fig. \ref{fig:steady-state-solution})
\begin{equation}
    \Sigma(r) \propto r^{-1+\xi},
    \label{eq:steady-state}
\end{equation}
Eq. (\ref{eq:master-eq-final}) yields to the quadratic equation
\begin{equation}
   \xi^2+\frac{1}{2} (1+\psi) \xi - \frac{\psi}{\revmnras{4}(\lambda-1)} =0
    \label{eq:xi-equation}
\end{equation}
that gives
\begin{equation}
\xi = \frac{1}{4} \left( \psi+1 \right) 
\left[ \sqrt{1+\frac{4 \psi }{(\lambda-1)(\psi+1)^2}}-1\right].
\label{eq:xi}
\end{equation}

The parameter $\xi$ appears as the mass ejection index
\begin{equation}
\xi \equiv \frac{d\ln{\Macc}}{d\ln{r}}
\label{eq:xi-Macc-def}
\end{equation}
that quantifies the local mass-loss rate relative to the local accretion rate as defined by \citet{1995A&A...295..807F}. In the presence of a wind, $\xi$ is non-vanishing and $\dot{M}_{acc}(r)$ varies across the disc due to the mass loss (see Fig. \ref{fig:steady-state-solution} and next subsection). The higher $\xi$ is, the more mass is locally ejected to drive accretion and the flatter the surface density profile is. 

The mass ejection index, plotted in Fig. \ref{fig:xi}, depends on both $\lambda$ and $\psi$. For $\lambda > 3/2$, the surface density profile is always decreasing with radius as $\xi<1$. $\xi$ is larger for lower $\lambda$ as winds of low $\lambda$ drive disc accretion with higher mass-loss rates. In the limit of pure wind accretion ($\psi = +\infty$), we recover the standard relation $\xi= 1/[2(\lambda-1)]$ \citep{1997A&A...319..340F}. For a finite value of $\psi$, the ejection index is smaller than this value as a non-vanishing fraction of mass is accreted via turbulence, without any mass loss. Specifically, for $\lambda \gtrsim 2$,
\begin{equation}
\xi \simeq  \frac{1}{2(\lambda-1)} \frac{\psi}{\psi+1},
\end{equation}
which translates the fact that the mass ejection index is about that of pure wind accretion reduced by the fraction of angular momentum effectively extracted by the wind. In other words, as the wind torque increases with respect to the turbulent torque, the surface density profile gets flatter.

We note that for $\alpha$-parameters and a temperature with arbitrary power-law dependence on radius, the accretion velocity varies with radius, resulting in a surface density that scales as $\Sigma \propto r^{ \xi -\gamma}$,
where $c_S^2 \tilde{\alpha} \propto r^{\gamma-3/2}$ and $\psi$ constant. For $\gamma < \xi$, the surface density increases with radius, a situation that could describe a cavity opening in the disc. For conciseness, we focus here on the case $\gamma = 1$ and $\lambda > 3/2$.

\subsubsection{Accretion and mass loss rate}
\label{subsubsec:steady-acc-loss}

\begin{figure}
	\includegraphics[width=\columnwidth]{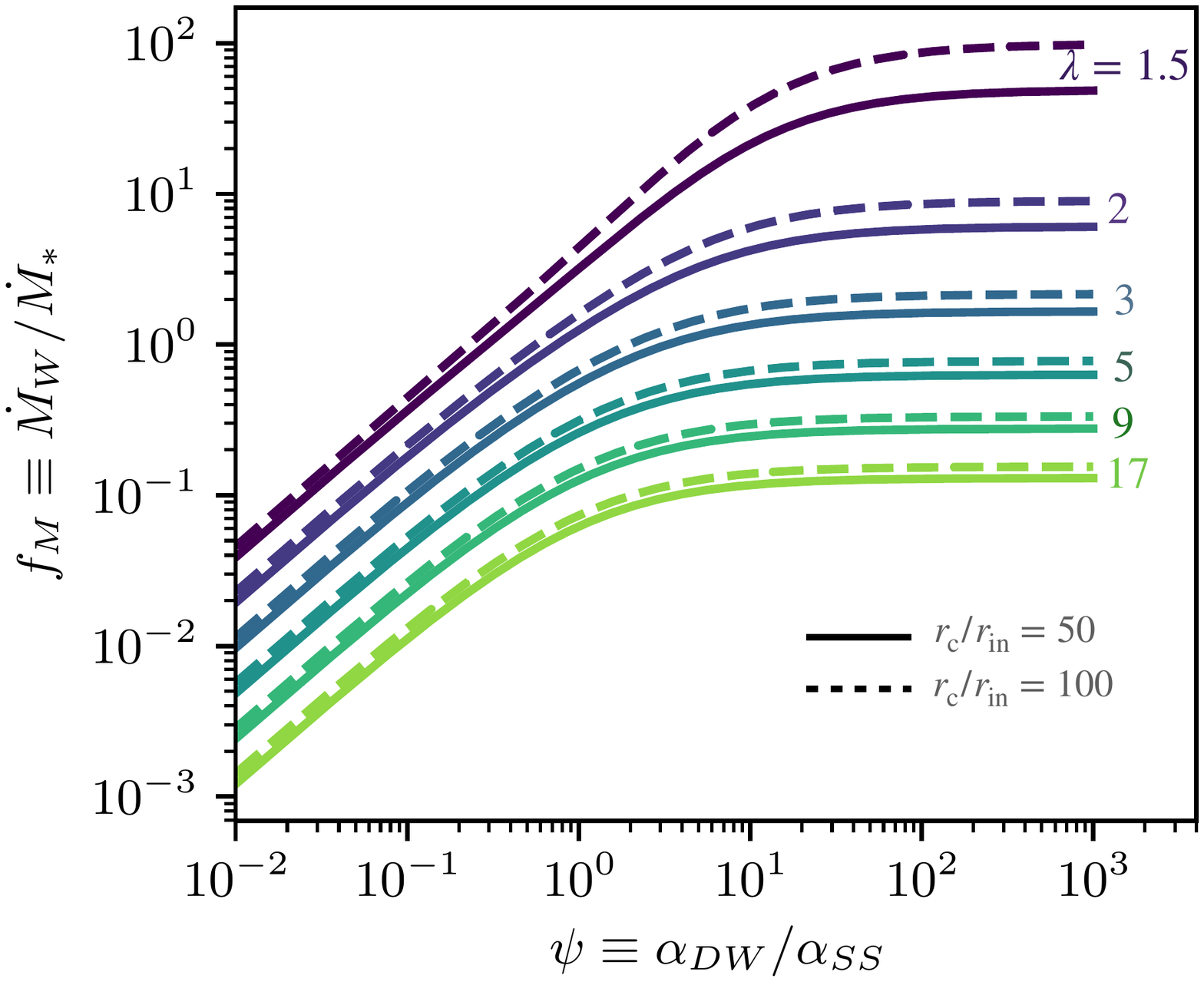}
    \caption{Global mass ejection-to-accretion ratio $f_M$ corresponding to the fraction of mass that is ejected in the wind compared to that accreted to the central star. As show in Eq. (\ref{eq:xi-def-steady-state}), $f_M$ depends on $\psi$, $\lambda$ (via $\xi$), and on the radial extent of the disc $\rc/\rin$, where $r_c$ denotes either an arbitrary radius in the steady-state solution (see Fig. \ref{fig:steady-state-solution}), or the characteristic disc radius of the self-similar solution (see Fig. \ref{fig:schematic-similarity}). 
    }
    \label{fig:fM} 
\end{figure}

Due to the local ejection of mass, the local accretion rate $\Macc(r)$ varies across the "leaky" disc (see Fig. \ref{fig:steady-state-solution}). 
With this notations, Eq. (\ref{eq:xi-Macc-def}) yields a local accretion rate of
\begin{equation}
 \dot{M}_{acc}(r) = \dot{M}_* \left(\frac{r}{r_{in}} \right)^{\xi},
 \label{eq:Macc-local}
\end{equation}
where $r_{in}$ denotes the inner radius of the disc and $\dot{M}_*$ denotes the stellar accretion rate. As discussed in Sec. \ref{subsubsec:simplifying-assumptions}, $r_{in}$ is not necessarily the inner radius of the disc which is about $0.05$~au in T Tauri discs, but can be considered as the inner radius of the wind launching region.

By mass conservation, the cumulative mass loss rate integrated between $r_{in}$ and an arbitrary radius $r$ is
\begin{equation}
 \dot{M}_{W}(r) = \dot{M}_* \left[\left(\frac{r}{r_{in}} \right)^{\xi}-1 \right],  
 \label{eq:MW-local}
\end{equation}
Figure \ref{fig:steady-state-solution} shows that the mass loss rate increases as a power-law of the radius at large distance. All in all, the mass loss rate can be much higher than the stellar accretion rate, resulting in a steep drop in the accretion rate toward $\rin$. 

It is then convenient to define the dimensionless mass ejection-to-accretion ratio estimated at a radius $r_c$ as \citep{2020A&A...640A..82T}
\begin{equation}
f_M(r_c/r_{in}) \equiv \frac{\dot{M}_{W}}{\dot{M}_{*}} = \left( r_c/r_{in} \right)^{\xi}-1.
\label{eq:xi-def-steady-state}
\end{equation}
This is considered to be a global parameter that quantifies the mass loss-rate over a disc extending from $r_{in}$ out to $r_c$. The higher $f_M$ is, the less mass passing though the disc at $r_c$ makes it to the growing star and the more mass is carried away by the wind. 

Figure \ref{fig:fM} shows that $f_M$ increases with increasing $\psi$ as the fraction of the accretion flow mediated by the vertical torque increases. $f_M$ increases with decreasing $\lambda$ as more mass is then launched to extract the amount of angular momentum required to drive wind accretion. 

Quantitatively, when the turbulent torque dominates ($\psi \ll 1$), the mass loss rate is negligible. \revmnras{In this limit, a Taylor expansion of Eq. (\ref{eq:xi-def-steady-state}) ($\xi \ll 1$) gives}
\begin{equation}
    f_M(r_c/r_{in}) \simeq \frac{1}{2(\lambda-1)} \psi \ln(\rc/r_{\text{in}}).
    \label{eq:fM-low-limit}
\end{equation}
$f_M$ is then proportional to $\psi$ (see Fig. \ref{fig:fM}). When the wind torque dominates ($\psi \gg 1$), $f_M$ converges to the pure wind value of
\begin{equation}
    \revmnras{f_M(r_c/r_{in}) = (\rc/r_{\text{in}})^{\frac{1}{2(\lambda-1)}}-1}
    \label{eq:fM-wind-limit}
\end{equation}
which can be either large or small, depending on the radial extent of the wind $\rc/r_{\text{in}}$ and on $\lambda$ (see Fig. \ref{fig:fM}). 
In particular, in the wind dominated regime, the mass loss rate is larger than the stellar accretion rate (i.e. $f_M \ge 1$) for \begin{equation}
\lambda \le 1+ \frac{\ln\left(r_{\rm{c}}/\rin\right)}{2 \ln(2)},
\end{equation}
which corresponds to $\lambda \le 3.8$ for $\rc/r_{\text{in}}=50$.

\subsection{General form of the self-similar solution}
\label{subsec:general-form-solution}
\begin{figure}
	\includegraphics[width=\columnwidth]{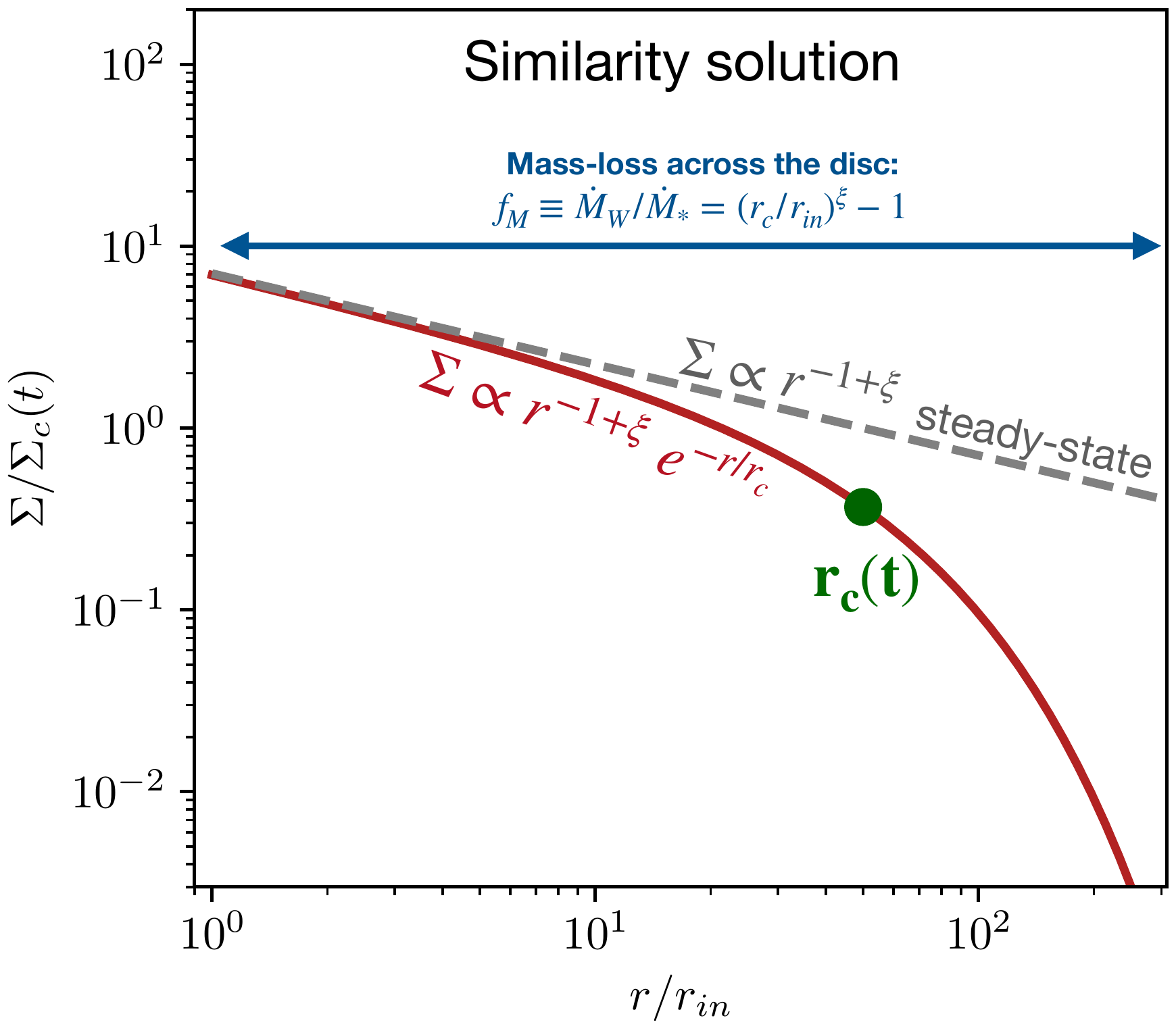}
    \caption{\rev{Surface density profile of the self-similar solution. The core is a power-law that corresponds to the steady-state solution (see Fig. \ref{fig:steady-state-solution}) and is tapered by an exponential cutoff at large distance.} The tapering is controlled by $r_c(t)$, the characteristic disc radius. During the evolution of the disc, the functional form of the surface density pictured here remains unchanged but $\Sigmac(t)$ and $r_c(t)$ varies differently for each class of solution (see Table \ref{table:formula}). At any time, the total mass loss-rate is quantified by $f_M$, which depends on the radial extent of the disc $r_{c}/r_{in}$, $\lambda$, and $\psi$ as shown in Fig. \ref{fig:fM}.}
    \label{fig:schematic-similarity} 
\end{figure}

\subsubsection{Ansatz}
The steady-state solution presented in Sec. \ref{subsec:steady-solution} assumes a disc of infinite size. For a finite disc size, i.e. a surface density that eventually decreases sharply with radius, the surface density profile depends on time as material coming from the outer regions is advected toward the inner regions. 
\citet{1974MNRAS.168..603L} derived analytical solutions for a turbulent disc that is, \rev{in the case of constant $\alphaSS$ and $T(r) \propto r^{-1/2}$}, of the form:
\begin{equation}
    \Sigma(r,t) = \Sigmac \left(\frac{r}{r_c}\right)^{-1} e^{-r/r_{c}},
\end{equation}
where $\Sigmac$ and $r_c$ are functions of time only. In particular, the core of the surface density profile is a power-law that corresponds to the steady-state solution (Eq. (\ref{eq:steady-turb})).

Inspired by this solution and by the steady-state solutions presented above (Eq. (\ref{eq:steady-state})), we find exact solutions of Eq. (\ref{eq:master-eq-final}) using the ansatz
\begin{equation}
\Sigma(r,t) = \Sigma_{c}(t) \left( \frac{r}{r_{\rm{c}}(t)}\right)^{-1+\xi} e^{-r/r_{\rm{c}}(t)},
\label{eq:sigma-ansatz}
\end{equation}
where $r_{\rm{c}}(t)$ is the characteristic disc radius, $\xi$ is the mass ejection index defined in Eq. (\ref{eq:xi}), and $\Sigmac(t)$ is a function of time only. \rev{As shown in Fig. \ref{fig:schematic-similarity}, at any time, the surface density profile is power-law in the inner region tapped by an exponential at large radius. The tapering is controlled by a characteristic disc radius $\rc(t)$  which defines the disc size.}

The quantity $\Sigmac(t)$ in Eq. (\ref{eq:sigma-ansatz}) can also be expressed in terms of the disc mass as\footnote{The exact normalization of the surface density profile includes a multiplicative factor $\Sigmac = M_D / (2\pi r_c^2 \Gamma(\xi+1))$, where $\Gamma$ is the gamma function. For $\lambda > 3/2$ one finds $\Gamma(\xi+1)\simeq 1$ and in the following, we adopt $\Gamma(\xi+1) = 1$ (see Appendix \ref{app:disc-mass}).}
\begin{equation}
    \Sigmac(t) = \frac{M_D(t)}{2\pi r_{\rm{c}}(t)^2},
    \label{eq:MD-Sigma0}
\end{equation}
for $r_c \gg r_{in}$. The self-similar ansatz can also be extended to the case \revmnras{of $T(r)$, $\alphaDW$, and $\alphaSS$ with arbitrary} power-law dependence on radius (Appendix \ref{app:gamma-solutions}). 


\subsubsection{Classes of solution}
\label{subsec:classes-solution}
The ansatz (\ref{eq:sigma-ansatz}) provides exact solutions of Eq. (\ref{eq:master-eq-final}) in the case of $\alphaSS$ and $\lambda$ constant in time. Regarding the $\alphaDW$ parameter, we explore two classes of solution.
\begin{enumerate}
    \item \textbf{Hybrid solutions} (Sec. \ref{subsec:solution-cste-alphaDW}), for which $\alphaDW$ is constant in time. This is the simplest solution that highlights essential features of the wind-driven accretion compared with turbulent accretion. The parameter $\psi \equiv \alphaDW/\alphaSS$ is the key parameter of these solutions.
    
    \item \textbf{$\Sigma_{\text{c}}$-dependent wind torque solutions} (Sec. \ref{subsec:solution-alphaDW-sigma}), for which $\alphaDW$ is constant across the disc but varies implicitly with time as \textit{$\alphaDW(t) \propto \Sigmac(t)^{-\omega} $}, where $\omega$ is a free parameter and $\Sigmac$ is defined in Eq. (\ref{eq:sigma-ansatz}). These solutions describe the unknown evolution of magnetic field strength (see Sec. \ref{subsubsec:simplifying-assumptions}). For simplicity, we adopt $\alphaSS=0$. As shown below, the latter assumption leads to a constant $\rc(t)$. Since $M_D = 2\pi \Sigmac \rc(t)^2$, it follows that these solutions can also be considered as solutions for which $\alphaDW \propto M_D(t)^{-\omega}$. $\omega=1$ can be interpreted as a case for which the magnetic field strength is constant over time. The case $\omega>1$ is not considered as it corresponds to a magnetic field strength increasing with time. A full dispersal of the disc at finite time is predicted for $\omega > 0$. In the following, we focus on $0<\omega < 1$.
\end{enumerate}

We further define \textbf{a fiducial solution} for which $\alphaSS=0$ ($\psi = \infty$) and $\omega=0$. This corresponds to the limit of the two classes of solution, that is a solution without turbulent accretion and with $\alphaDW$ constant in time. \rev{We leave for future papers the general case for which $\alphaDW$ varies with time ($\omega \neq 0$) and $\alphaSS \neq 0$ (finite value of $\psi$) as it requires the use of numerical solutions.}

\subsubsection{Parameters of the self-similar solutions}

\begin{table}
\caption{Parameters of the self-similar solutions. The first block gives the free parameters of the solution and the second block gives additional disc parameters that are set by the free parameters.} 
\begin{tabular}{| c | | c | | c}
\hline
Parameter& Description & Ref. \\
\hline                       
$\lambda$ & magnetic lever arm parameter & \\ 
$\psi$ & wind-to-turbulent $\alpha$ ratio  & Eq. (\ref{eq:psi-def}) \\
$\omega$ & power-law index $\alphaDW(\Sigmac)$ & \\
$\tacc$ & initial accretion timescale & Eq. (\ref{eq:tacc-def}) \\
$r_c(0)$ & initial disc characteristic size \\
$r_c(0)/\rin$ & initial radial extend of the wind \\
$M_0$ & initial disc mass \\
\hline
$\xi$ & mass ejection index & Eq. (\ref{eq:xi}) \\
$f_{M,0}$ & initial mass ejection-to-accretion ratio & Eq. (\ref{eq:fM0-def}) \\
$\dot{M}_{*,0}$ & initial stellar accretion rate & Eq. (\ref{eq:Mdot-initi})\\
\hline
\hline
\end{tabular}
\label{table:param}
\end{table}

The time evolution of the surface density is controlled by the initial accretion timescale\footnote{We note that \citet{2020MNRAS.498.2845S} call 'accretion timescale' the ratio $M_D/\dot{M}_*$. In our work, the latter quantity is called 'disc lifetime' and is defined in Sec. (\ref{subsec:disc-lifetime}).}
\begin{equation}
\tacc \equiv \frac{r_c(t=0)}{3 \epsilon_c c_{s,c} \tilde{\alpha}(t=0)}, 
\label{eq:tacc-def}
\end{equation}
where $\epsilon_c$ is the disc aspect ratio and $c_{s,c}$ is the sound speed at $r = r_c(t=0)$. The initial accretion timescale corresponds to the time that would be required to accrete a fluid particle located initially at $\rc/2$~to the inner region of the disc with an accretion velocity equal to its initial value. With this definition, the initial viscous timescale writes
\begin{equation}
\begin{split}
t_{\nu,0} = (1+\psi) \tacc.
\label{eq:tnu0-def}
\end{split}
\end{equation}

The accretion rate and the total mass loss rate depends also on the initial radial extent of the disc $r_{c}(0)/r_{in}$. We quantify the effect of the mass loss rate using the initial mass ejection-to-accretion ratio that writes (see Appendix \ref{app:fM})
\begin{equation}
f_{M,0} = \left( r_{c}(0)/r_{in} \right)^{\xi}-1.
\label{eq:fM0-def}
\end{equation}
The main advantage of using $f_{M}$ \revmnras{is that it is} an observational quantity that does not depend on the disc mass.
Interestingly, at any time, the mass ejection-to-accretion ratio of the self-similar solution simply corresponds to that of the steady state solution evaluated at $r=r_c(t)$ (see Appendix \ref{app:fM}):
\begin{equation}
f_{M}(t) = \left( r_{c}(t)/r_{in} \right)^{\xi}-1.
\label{eq:fM-def}
\end{equation}
Its dependency on $r_c/r_{in}$, $\psi$, and $\lambda$ is given in Fig. \ref{fig:fM} and discussed above (see Sec. \ref{subsubsec:steady-acc-loss}). In the following, we fix $r_c(0)/r_{in} =50$ so that $f_{M,0}$ is not considered as a free parameter as it is set by the values of $\psi$ and $\lambda$ via the mass ejection index $\xi$ (Fig. \ref{fig:xi}).

In other words, the self-similar solutions presented in this work are controlled by 7 parameters, namely (see summary in Table \ref{table:param}):
\begin{itemize}
    \item $\lambda$, the magnetic lever arm parameter, that has typical values of about $\lambda \simeq 2-5$ (see Sec. \ref{subsec:obs-lambda})
    \item $\psi$, the ratio between the wind torque and the turbulent torque, which is explored in the first class of solution,
    \item $\omega$, the power-law index of $\alphaDW$ with \revmnras{$\Sigmac$}, which is explored in the second class of solution,
    \item $\tacc$, the initial accretion timescale, which is related to the disc aspect ratio $\epsilon$, the initial characteristic disc radius, and the total torque, $\tilde{\alpha}$ (see Eq. \ref{eq:tacc-def}), and has typical values of $\tacc \simeq 1~$Myr (see Sec. \ref{subsec:disc-lifetime-tacc}), 
    \item $r_c(0)$, the initial characteristic disc radius, that has typical values of $r_{\text{c}} \simeq 20-100~$au,
    \item $r_c(0)/\rin$, the radial extent of the wind launching region set to 50 (see discussion about the value of $\rin$ in Sec. \ref{subsubsec:simplifying-assumptions}),
    \item $M_0$, the initial disc mass, ranging typically between $10^{-3}-10^{-2}~M_{\odot}$ \citep[from the masses of Class I discs,][]{2020A&A...640A..19T,2020ApJ...890..130T}.
\end{itemize}
\revmnras{The analytical solutions for the hybrid and $\Sigmac$-dependent wind torque solutions are derived in Sec. \ref{subsec:solution-cste-alphaDW} and \ref{subsec:solution-alphaDW-sigma} respectively and summarized in Table \ref{table:formula}. The evolution of the surface density profile is shown in Fig. \ref{fig:sigma} and discussed below.}
\begin{table*}
\caption{Disc quantities for the two classes of solution.}              
\centering                                      
\begin{tabular}{| c || c c c c|}          
\hline                       
Quantity & $\Sigmac(t)/\Sigmac(0)$ & $r_{\rm{c}}(t)/r_{\rm{c}}(0)$ & $M_D(t)/M_0$ & $\dot{M}_*(t)/\dot{M}_{*,0}$ \\
\hline 
\hline 
Fiducial ($\psi=\infty$, $\omega=0$) & $e^{-t/\revmnras{2}\tacc}$ & $1$ &  $e^{-t/\revmnras{2}\tacc}$ & $e^{-t/\revmnras{2}\tacc}$ \\
\hline 
Hybrid ($\omega =0$) & $\left( 1 + \frac{t}{(1+\psi) \tacc} \right)^{-\frac{1}{2}(\psi+2\xi+5)} $ & $\left(1+\frac{t}{(1+\psi) \tacc}\right)$ & $\left( 1 + \frac{t}{(1+\psi) \tacc} \right)^{-\frac{1}{2}(\psi+2\xi+1)} $ & $\left( 1+\frac{t}{(1+\psi)\tacc}\right)^{-\frac{1}{2} (\psi+4\xi+3)}  $ \\
\hline
$\Sigmac$-dep. $\alphaDW$  ($\psi =\infty$) &  $\left( 1-\frac{\omega}{2 \tacc} t  \right)^{1/\omega}$ & $1$  & $\left( 1-\frac{\omega}{2 \tacc} t \right)^{1/\omega}$ & $\left( 1-\frac{\omega}{2 \tacc} t  \right)^{-1+1/\omega}$ \\
\hline
\end{tabular}
\label{table:formula}
\end{table*}



\subsection{Hybrid solutions (constant $\alpha$ parameters)}
\label{subsec:solution-cste-alphaDW}

\subsubsection{Solution}
Assuming that $\alphaSS$ and $\alphaDW$ are constant in time and space, the master equation Eq. (\ref{eq:master-eq-final}) can be written using the dimensionless coordinates $\tilde{r} = r/r_{c}(0)$ and $\tilde{t}= t/t_{\nu,0}$ as
\begin{equation}
\begin{split}
\frac{\partial \Sigma(\tilde{r},\tilde{t})}{\partial \tilde{t}} = &
\frac{1}{\tilde{r}}\frac{\partial}{\partial \tilde{r}} \left( \tilde{r}^{1/2} \frac{\partial}{\partial \tilde{r}} (\tilde{r}^{3/2} \Sigma(\tilde{r},\tilde{t})) \right) \\
+ & \frac{\psi}{2 \tilde{r}} \frac{\partial}{\partial \tilde{r}} \left\{\tilde{r} \Sigma(\tilde{r},\tilde{t}) \right\} -\frac{\psi}{4 (\lambda-1) \tilde{r}}  \Sigma(\tilde{r},\tilde{t})~~.
\end{split}
  \label{eq:master-eq-appendix}
\end{equation}

Rewriting the self-similar ansatz (\ref{eq:sigma-ansatz}) as $\Sigma(\tilde{r},\tilde{t}) = A(\tilde{t}) \tilde{r}^{-1+\xi} e^{-\tilde{r}/\tilde{\rc}(\tilde{t})}$, Eq. (\ref{eq:master-eq-appendix}) yields a system of 2 ordinary differential equations
\begin{equation}
\begin{split}
  & \dot{\tilde{r}}_{\rm{c}}(\tilde{t})  = 1, \\
  & \frac{\dot{A}(\tilde{t})}{A(\tilde{t})} = -\frac{1}{\tilde{r}_{\rm{c}}(\tilde{t})} \left( \frac{3}{2} + 2\xi + \frac{\psi}{2} \right).
\end{split}
\end{equation}
The solution of this system in dimensional form is
\begin{equation}
\begin{split}
   & r_{\rm{c}}(t)  = r_{\rm{c}}(0) \left(1 + \frac{t}{(1+\psi) \tacc} \right), \\
    & \Sigmac(t) = \Sigmac(0) \left(1 + \frac{t}{(1+\psi) \tacc} \right)^{-(\frac{5}{2} + \xi + \frac{\psi}{2} )}.
\end{split}
\label{eq:solution-case1}
\end{equation}

%
\begin{figure*}
	\includegraphics[width=\linewidth]{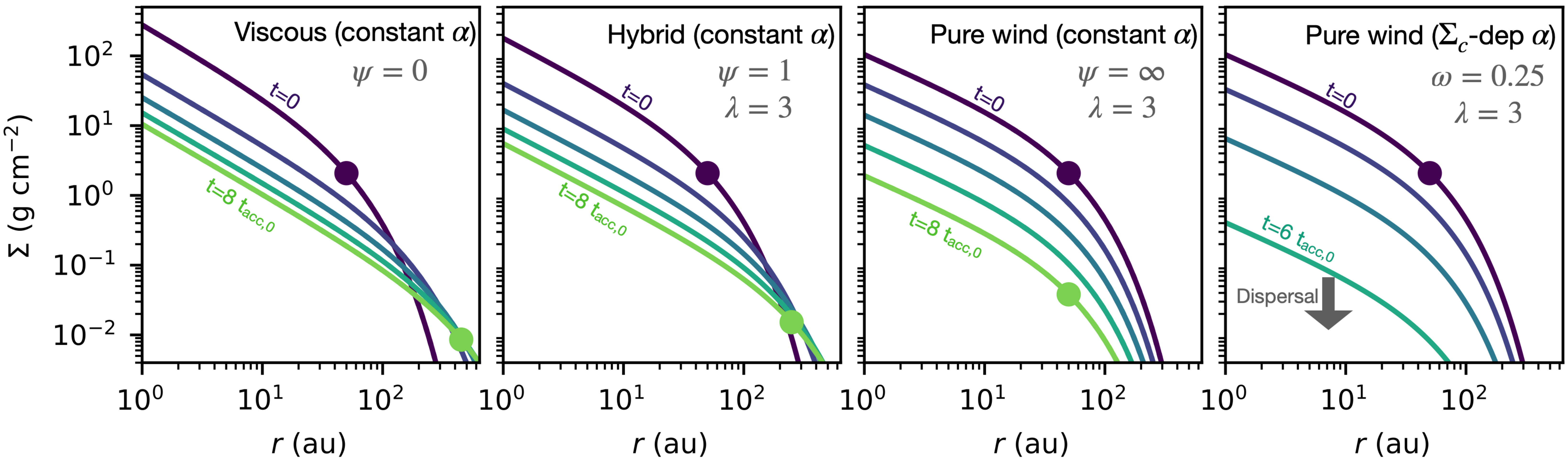}
    \caption{\revmnras{Examples of the evolution of the surface density for the analytical solutions presented in this work. The first three left panels are for constant $\alpha$ solutions (see Sec. \ref{subsec:solution-cste-alphaDW}) in the case of viscous ($\psi=0$), hybrid ($\psi=1$), and pure wind-driven ($\psi=\infty$) accretion. The rightmost panel is for a $\Sigmac$-dependent wind torque solution (see Sec. \ref{subsec:solution-alphaDW-sigma}) with $\omega=0.25$. At initial time, the discs share the same initial characteristic radius of $r_{\rm{c}}(0)=50~$au and disc mass of $M_D=10^{-2} M_{\odot}$. The time evolution of the surface density is controlled by the initial accretion timescale $\tacc$ (see definition in Eq. \ref{eq:tacc-def}).} Each line corresponds to a different age ranging from $t=0$ to $8 \tacc$ with steps of $2 \tacc$ (from dark blue to green). \revmnras{In the $\Sigmac$-dependent wind torque solution (rightmost panel), the disc is dispersed at $t=8\tacc$.} The characteristic radius $r_c(t)$, outlined by dots, is constant in the pure wind cases, whereas it increases in the viscous cases. The power-law index in the inner tens of au is flatter in the wind cases due to the mass-loss accompanying the accretion flow (see Eq. (\ref{eq:steady-state})).}
    \label{fig:sigma}
\end{figure*}

Examples of the time evolution of the surface density $\Sigma(r,t)$ for various values of $\psi$ are shown in Fig. \ref{fig:sigma}, for the same initial mass, radius, and accretion timescale. Starting from similar surface density profiles, the surface densities differ rapidly over less than an accretion time-scale. This results in a significantly different evolution of the global disc properties $r_c(t)$, $M_D(t)$, $\dot{M}_*(t)$, and $\dot{M}_W(t)$ that are studied below.

\subsubsection{Disc radius}

The location of the disc characteristic radius $\rc(t)$ is outlined in Fig. \ref{fig:sigma} by a circular marker.
The time evolution of $\rc(t)$ depends on the dominant accretion process quantified by $\psi$. In the pure viscous case ($\psi=0$), the disc spreads as angular momentum it transported radially. In contrast, in the pure wind case (fiducial solution, $\psi=+\infty$), the disc characteristic radius remains constant, as all the angular momentum is extracted vertically, without any need for disc spreading. 
 
In the hybrid case, the disc spreads, though at a slower rate than in the pure viscous case. The disc characteristic radius provided in Eq. (\ref{eq:solution-case1}) can be rewritten as
\begin{equation}
r_{\rm{c}}(t) = r_{\rm{c}}(0) \left(1 + \frac{t}{t_{\nu,0}} \right),
\label{eq:disc-radius}
\end{equation}
where $t_{\nu,0}$ is the initial viscous timescale \revmnras{(see Eq. (\ref{eq:tnu0-def}))}. Interestingly, we find that the disc spreading is not affected by the presence of the wind and we simply recover the classical relation of the pure viscous accretion. In particular, for $t \gg t_{\nu,0}$, the disc size increases linearly with time. However, for a fixed accretion timescale $\tacc$, the disc spreading timescale increases with increasing wind torque since $t_{\nu,0} = (1+\psi) \tacc$.

\subsubsection{Disc mass}

\begin{figure}
	\includegraphics[width=\columnwidth]{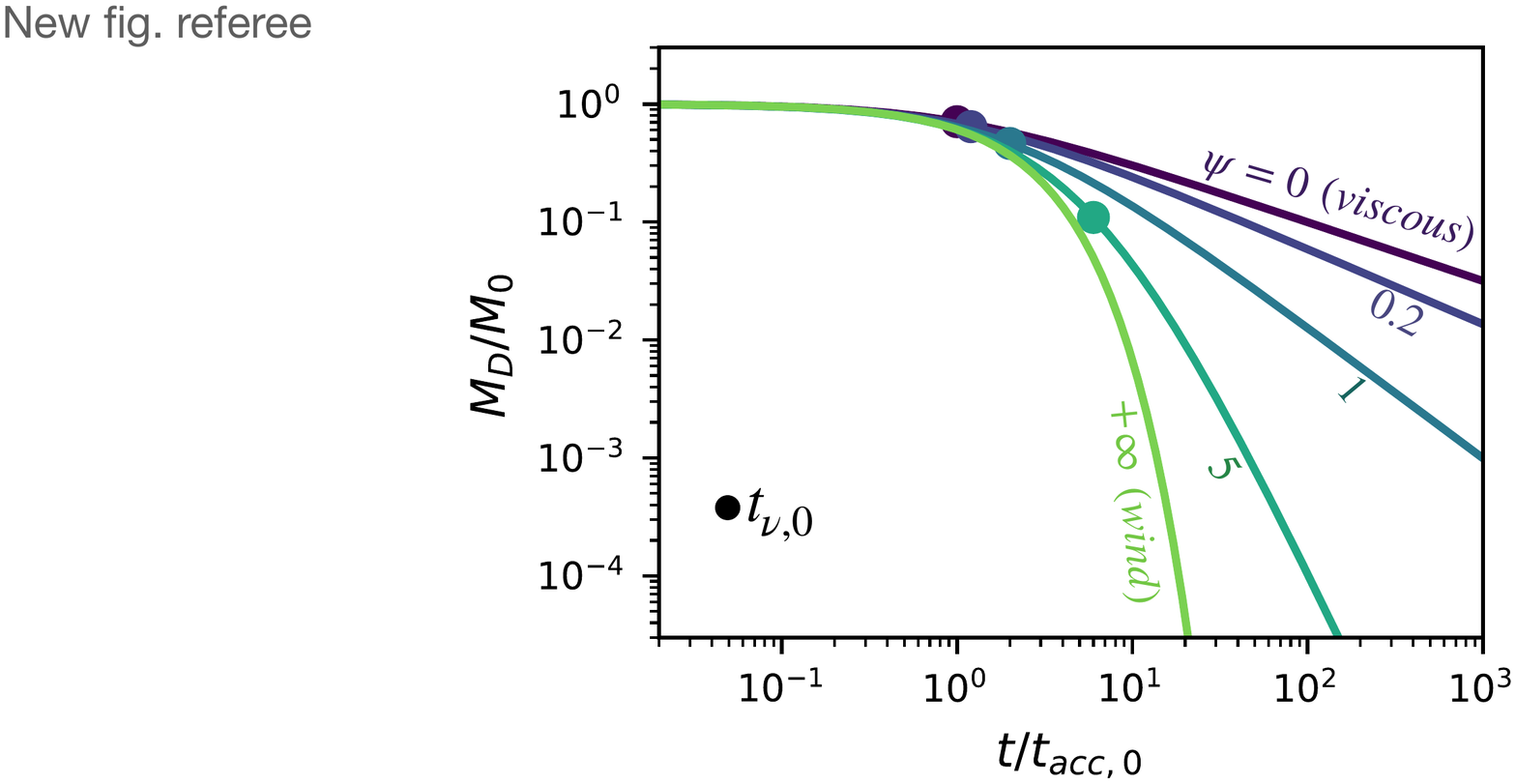}
    \caption{Time evolution of the disc mass \revmnras{for the hybrid solutions,} depending on the relative contribution of the vertical to the radial torque quantified by $\psi$. In all cases $\lambda= 3$ so that $\xi$ varies between $0$ and $0.25$. The value of the viscous accretion timescale $t_{\nu,0} = (1+\psi)\tacc$ is indicated with circle along each curve.}
    \label{fig:disc-mass} 
\end{figure}

The evolution of the disc mass depends also critically on the driving accretion process.
The evolution of the disc mass obtained from Eq. (\ref{eq:MD-Sigma0}) and (\ref{eq:solution-case1}) is 
\begin{equation}
M_D(t) = M_0 \left( 1 + \frac{t}{(1+\psi) \tacc} \right)^{-\frac{1}{2}(\psi+2\xi+1)},
\label{eq:disc-mass}
\end{equation}
and shown in Fig. \ref{fig:disc-mass}. In the viscous case ($\psi=0$), the disc mass decreases slowly with time with the classical scaling $M_D(t)\propto 1/\sqrt{t}$ for $t\gg t_{\nu,0}$. This is ultimately due to the viscous spreading of the disc. As the gas in the bulk part of the disc is advected at constant velocity, the instantaneous viscous timescale (i.e. the time required to advect the gas located at $\rc(t)/2$ to the inner disc) increases as $t_{\nu,0}(t) \propto t$ due to the increase in $\rc(t)$. This prevents the disc from rapidly draining all its material onto the star. 

In the fiducial wind-driven case, the disc mass drops by more than two orders of magnitudes after $t=10 \tacc$ (see Fig. \ref{fig:disc-mass}). Quantitatively, a Taylor expansion of Eq. (\ref{eq:disc-mass}) in the case $\psi \rightarrow +\infty$ shows that the evolution of the disc mass follows an exponential decay with
\begin{equation}
M_D(t) = M_0 e^{-\frac{t}{2\tacc}}.
\label{eq:disc-mass-fiducial}
\end{equation}
This behaviour, which contrasts with the viscous case, is due to the absence of disc spreading. Since the outer radius remains constant, the bulk part of the disc located initially within $r \lesssim r_c(0)$~is drained after $t \simeq 2 \tacc$. \rev{The evolution of the disc mass does not depend on the fraction of mass lost in the wind quantified by $f_M$. Therefore neither the value of $\lambda$ nor $r_c/r_{in}$ impacts the evolution of the disc mass (see Eq. (\ref{eq:fM0-def})).}

In the hybrid case (see Fig. \ref{fig:disc-mass}), the evolution of the disc mass is bracketed between that of the exponentially decaying pure wind case ($\psi = \infty$) and that of the slowly evolving pure viscous case ($\psi = 0$), and appears to be sensitive to the value of $\psi$. For $t \gg t_{\nu,0} = (1+\psi)\tacc$, the disc mass exhibits a power-law dependence on time
\begin{equation}
M_D(t) \simeq M_0 \left(\frac{t}{t_{\nu,0}}\right)^{-\frac{1}{2}(\psi+2\xi+1)}
\end{equation}
as in the viscous case, but with a power-law index that is increased by $\frac{1}{2}(\psi+2\xi)$. This is a typical hybrid behavior: the viscous spreading prevents the rapid draining of the disc, but the wind accretion ensures a steeper decrease in the disc mass with time due to its ability to sustain high accretion rate.

Interestingly, in the more general case for which $c_S^2 \tilde{\alpha} \propto r^{\gamma-3/2}$, the disc mass scales as $(1+t/t_{\nu,0})^{-(1+2\xi+\psi)/(2(2-\gamma))}$ (see Appendix \ref{app:gamma-solutions}). Therefore, there is a degeneracy in the slope of the disc mass, between $\gamma$ and $\psi$. For example, a pure viscous case with a radial gradient of $\alphaSS  \propto r^{-1+\gamma}$ and a hybrid case with constant $\alpha$-parameters have the same slope for
\begin{equation}
 \psi = \frac{\gamma-1}{2-\gamma},
 \label{eq:equivalence}
\end{equation}
where we assumed $\xi \ll 1$ and a same power-law index of the temperature.

\begin{figure*}
	\includegraphics[width=\textwidth]{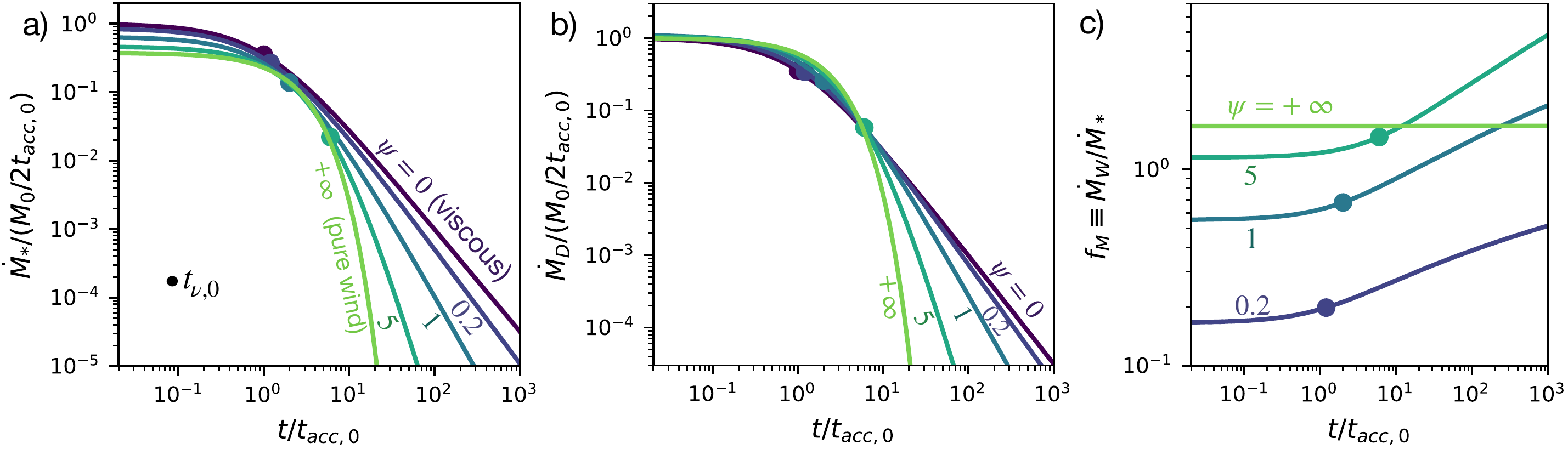}
    \caption{Evolution of the accretion rate and mass loss rate for the hybrid solutions. a) The stellar accretion rate, b) the time derivative of the disc mass, which is equal to the sum of the accretion rate and the mass loss rate, c) the mass ejection-to-accretion ratio $f_M=\dot{M}_W/\dot{M}_*$. In all cases we fixed $\lambda= 3$ and $r_{c}(0)/\rin = 50$.
    }
    \label{fig:mass-acc} 
\end{figure*}

\subsubsection{Accretion rate}

In the presence of a wind, the accretion rate is the time derivative of the disc mass reduced by the fraction mass that is effectively accreted onto the growing star:
\begin{equation}
    \dot{M}_{*}(t) = \frac{1}{1+f_M(t)} \dot{M}_D(t),
    \label{eq:dotM*-def}
\end{equation}
where $f_{M}(t)$ is the instantaneous mass ejection-to-accretion ratio given in Eq. (\ref{eq:fM-def}).
Therefore, at $t=0$, the stellar accretion rate is
\begin{equation}
\begin{split}
     \dot{M}_{*,0} & = \frac{\psi+1+2\xi}{\psi+1} \frac{M_0}{2 \tacc (1+f_{M,0})} \\ 
    & \simeq \frac{M_0}{2 \tacc (1+f_{M,0})} ~~ \text{for}~~\xi \ll 1 \revmnras{~~\textnormal{or}~~ \psi \gg 1}.   
\end{split}
    \label{eq:Mdot-initi}
\end{equation}
The effect of the mass loss rate can be seen in Fig. \ref{fig:mass-acc}-a, where the initial accretion rate is lower for higher values of $\psi$ as it corresponds to higher value of $f_{M,0}$. 

As time increases, the evolution of the accretion rate is driven by the evolution of $\dot{M}_D(t)$, but also by that of $f_M(t)$ (see Fig. \ref{fig:mass-acc}-b and c, respectively). In fact, as the disc spreads, $f_M(t)$ smoothly increases as the radial extent of the disc increases. However, the impact of the latter on the evolution of $\dot{M}_*(t)$ is generally negligible since fast viscous spreading is operating for $\psi \ll 1$, a regime where $f_M \ll 1$ (see Fig. \ref{fig:fM}). All in all, the stellar accretion rate is
\begin{equation}
\begin{split}
    &\dot{M}_{*}(t) =  \dot{M}_{*,0} \left( 1+\frac{t}{(1+\psi) \tacc}\right)^{-\frac{1}{2}(\psi+4\xi+3)} \\
    &\simeq  \frac{M_0}{2 \tacc (1+f_{M,0})}\left( 1+\frac{t}{(1+\psi) \tacc}\right)^{-\frac{1}{2}(\psi+3)}~~ \text{for}~\xi \ll 1 \revmnras{~\textnormal{or}~ \psi \gg 1}.
\end{split}
\label{eq:accretion-rate-2}
\end{equation}
In the pure viscous case ($\psi=0$) we recover the well-known scaling $\dot{M}_*(t)\propto t^{-3/2}$ for $t\gg t_{\nu,0}$. In the pure wind case $f_{M}(t)$ is constant (no disc spreading) and the accretion rate drops exponentially with time
\begin{equation}
\dot{M}_{*}(t) = \frac{M_0}{2 \tacc} \frac{1}{1+f_{M,0}} e^{-t/2\tacc}.
\label{eq:acc-rate-fiducial}
\end{equation}

When both viscous and wind torque are effective, the evolution of the accretion rate depends mostly on the ratio $\psi$. For $t \gg t_{\nu,0}$, $\dot{M}_{*}$ exhibits a power-law dependence on time, with $\dot{M}_* \propto t^{-\frac{1}{2}(\psi+3)}$. For the more general case of $c_s^2 \tilde{\alpha} \propto r^{-3/2+\gamma}$, we recover the degeneracy found for the disc mass between $\psi$ and $\gamma$.

\subsubsection{Mass loss rate}
By symmetry, the mass-loss rate of the wind is
\begin{equation}
\dot{M}_{w}(t) =\frac{f_M(t)}{1+f_M(t)} \dot{M}_D(t).
\label{eq:mass-loss-rate}
\end{equation}
For $f_M \ll 1$, $\dot{M}_{w}$ is simply $\dot{M}_D$ reduced by $f_M$. 
For $f_M \gg 1$, $\dot{M}_{w}$ is about $\dot{M}_D$ as most of the disc mass is lost in the wind rather than accreted onto the star.

In the pure wind-driven case, the mass loss rate follows an exponential decrease 
\begin{equation}
\dot{M}_{w}(t) = \frac{M_0}{ 2 \tacc} \frac{f_{M0}}{1+f_{M0}} e^{-\frac{t}{2\tacc}}.
\label{eq:mass-loss-rate-wind}
\end{equation}









\subsection{$\Sigmac$-dependent wind torque (time-dependent $\alphaDW$)}
\label{subsec:solution-alphaDW-sigma}

\subsubsection{Solution}

In this section, we assume that the turbulent torque vanishes ($\alphaSS=0$, i.e. $\psi=\infty$) and the wind torque writes $\alphaDW(t) \propto \Sigmac(t)^{-\omega}$. Injecting the self-similar ansatz (\ref{eq:sigma-ansatz}) in Eq. (\ref{eq:master-eq-final}) yields to a system of 2 ordinary differential equations
\begin{equation}
\begin{split}
\dot{r}_c(\tilde{t}) & = 0, \\
\dot{\Sigma}_c(\tilde{t}) & = -\frac{1}{2} {\Sigma}_c(\tilde{t})^{1-\omega},
\end{split}
\end{equation}
where we used the dimensionless time $\tilde{t}= t/\tacc$.

The disc characteristic radius is then constant as all the angular momentum is extracted vertically ($\alphaSS=0$). The surface density is controlled by
\begin{equation}
\Sigmac(t) = \Sigmac(0)\left( 1-\frac{\omega}{2\tacc}t \right)^{1/\omega}
\label{eq:solution-case2}.
\end{equation}

\revmnras{An example of the time evolution of the surface density for $\omega=0.25$ is shown in Fig. \ref{fig:sigma} (right most panel). The surface density profile keeps the same shape as in the constant $\alphaDW$ solution (in the pure wind case). In particular, the slope of the core of $\Sigma(r,t)$ (i.e., for $r\ll r_c$) remains unchanged with $\Sigma(r,t) \propto r^{-1+\xi}$. This is due to the fact that despite the increase in $\alphaDW(t)$ with time, $\xi$ remains constant since is depends only on $\lambda$, which is kept constant. However, the absolute value of the surface density drops faster in the $\Sigmac$-dependent wind torque solution, as $\alphaDW(t)$ increases with time. This is best seen in the evolution of the disc mass presented below.}


\subsubsection{Global disc quantities}
\begin{figure}
	\includegraphics[width=0.95\columnwidth]{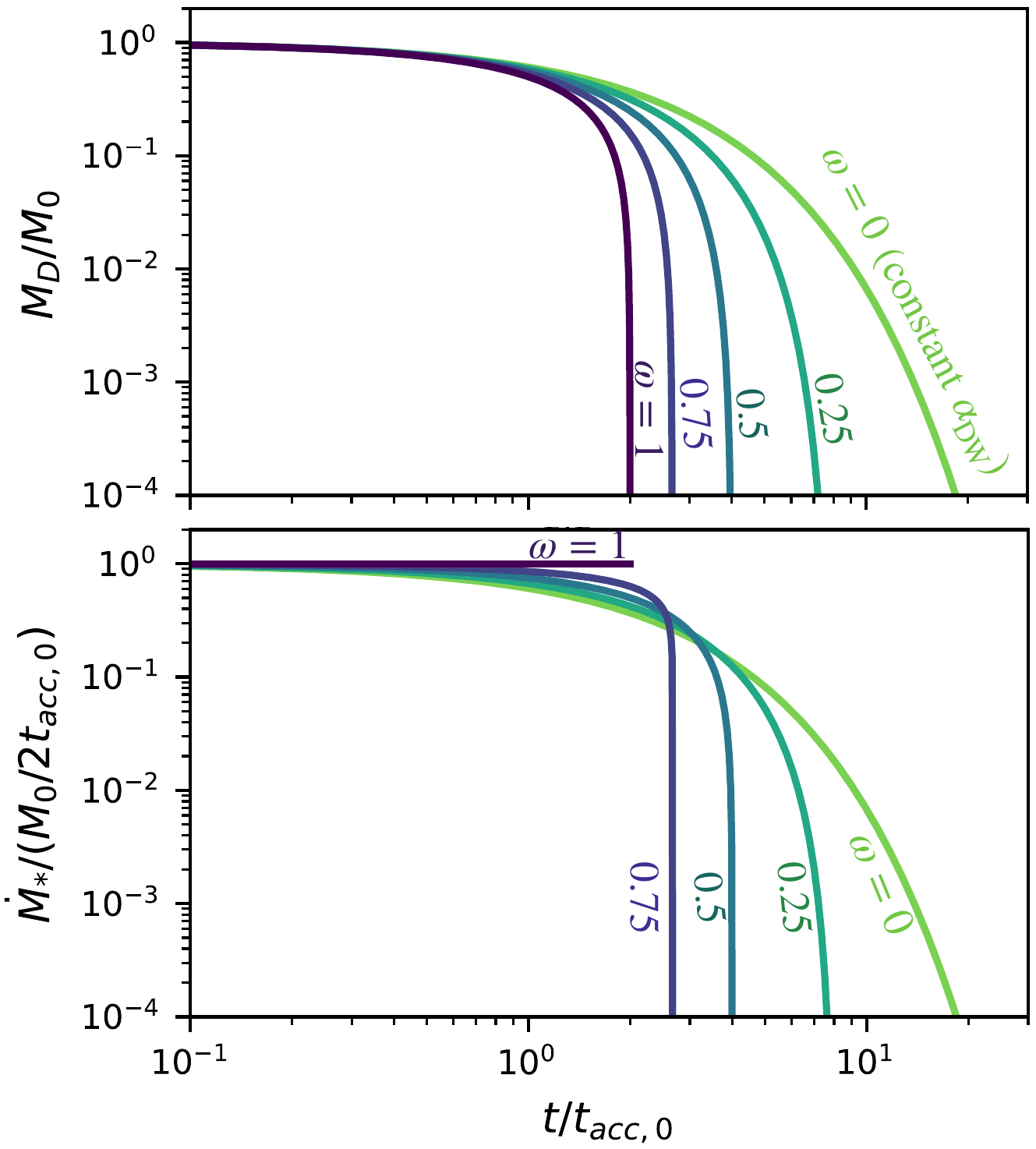}
    \caption{Evolution of the disc mass $M_D(t)$ (top panel) and the time derivative of disc mass $\dot{M}_D(t)$ (bottom panel) for the $\Sigma_{\text{c}}$-dependent wind torque solutions. The accretion rate $\dot{M}_*(t)$ and the wind mass loss rate $\dot{M}_W(t)$ are simply $\dot{M}_D(t)$ rescaled by $1/(1+f_{M,0})$ and by $f_{M,0}/(1+f_{M,0})$, respectively, where $f_{M,0}$ is defined in Eq. (\ref{eq:fM0-def}). 
    }
    \label{fig:mass-acc-disp} 
\end{figure}

From Eq. (\ref{eq:solution-case2}), the disc mass is
\begin{equation}
M_{D}(t) = M_0 \left( 1-\frac{\omega}{2 \tacc} t \right)^{1/\omega}. 
\label{eq:solution-case2-MD}
\end{equation}
The most striking feature of this new class of solution shown in Fig. \ref{fig:mass-acc-disp} (top panel) is the full dispersal of the disc after a finite time
\begin{equation}
t_{disp} \equiv \frac{2\tacc}{\omega},   
\end{equation}
for $\omega > 0$. This can be seen as a runaway accretion first described by \citet{2013ApJ...778L..14A}. In particular, injecting the solution (\ref{eq:solution-case2-MD}) in Eq. (\ref{eq:dotM*-def}) gives the stellar accretion rate
\begin{equation}
\dot{M}_{*}(t) = \frac{M_0}{2\tacc (1+f_{M,0})} \left( 1-\frac{\omega}{2 \tacc}t \right)^{-1+1/\omega},
\end{equation}
where the mass ejection-to-accretion ratio $f_M$, given by Eq. (\ref{eq:fM-def}), is constant in time as the disc radius \revmnras{and $\lambda$} remains constant in time. 

As the disc evolves, $\Sigmac(t)$ decreases, and $\alphaDW(t)$ increases. In the case $\omega=1$, the accretion rate, which is proportional to the product $\alphaDW(t) \Sigmac(t)$ is constant over time (Fig. \ref{fig:mass-acc-disp}, bottom). Therefore, the disc mass drops linearly with time and is dispersed after a time of $M_0/\dot{M}_* = 2M_0/\tacc$. This case is similar to that proposed by \citet[][see their Appendix C]{2017ApJ...847...31M}. 

For $0<\omega<1$, the increase in $\alphaDW(t)$ with decreasing $\Sigmac(t)$ is shallower. It follows that the decrease in the accretion rate is slower than that of the disc mass. Thus, for the same $\tacc$, the disc is dispersed at later time.

By symmetry, the mass loss rate of the wind is
\begin{equation}
\dot{M}_{w}(t) = \frac{M_0}{2\tacc} \frac{f_{M,0}}{1+f_{M,0}} \left( 1-\frac{\omega}{2\tacc}t \right)^{-1+1/\omega}.
\end{equation}
Because $f_{M}(t)$ is constant in time, $\dot{M}_{w}$ is simply the accretion rate multiplied by $f_{M,0}$ and follows the same functional form.

\revmnras{We note that since $\rc(t)$ is constant, a different radial profile of $T(r)$ or $\alphaDW$ will not change the time evolution of $M_D(t)$, $\dot{M}_*(t)$, and $\dot{M}_W(t)$. However, it will affect the radial profile of $\Sigma(r,t)$. For $\alphaDW c_s \propto r^{3/2-\gamma}$, one can show that $\Sigma \propto r^{\xi-\gamma} e^{-(r/r_c)^{2-\gamma}}$.}

\subsection{Solutions in the $\dot{M}_*-M_D$ plane}
\label{sec:MD-Macc}
\rev{With new surveys conducted in the visible and (sub)millimeter domains, it is now common to study accretion properties via the correlations between the accretion rates and the disc masses in populations of Class II sources of different ages \citep[e.g.,][]{2017ApJ...847...31M,2019A&A...631L...2M}. In this section, we therefore analyse the self-similar solutions presented above in the $\dot{M}_*-M_D$ plane.}


\subsubsection{Disc lifetime}
\label{subsec:disc-lifetime}
\begin{figure*}
	\includegraphics[width=2.\columnwidth]{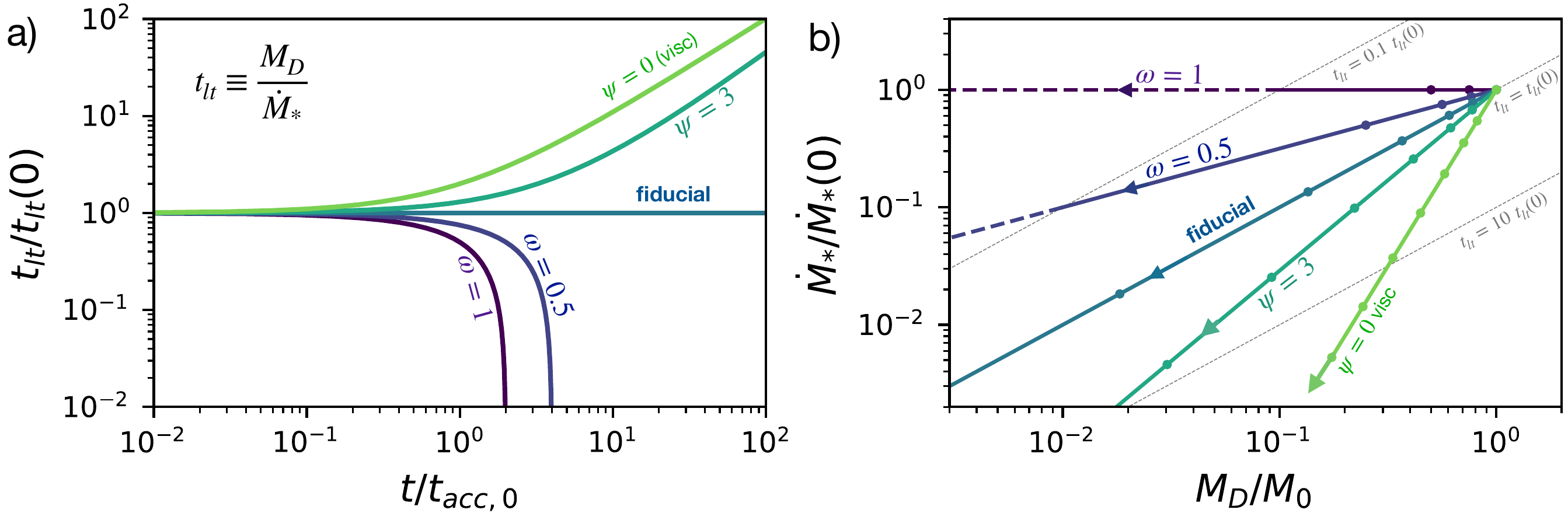}
    \caption{Joint evolution of accretion rate and disc mass. a) Evolution of the lifetime defined as $t_{lt}(t)\equiv M_D(t)/\dot{M}_*(t)$, depending on $\psi$ and $\omega$ for each class of solution. $t_{lt}(t)$ is normalized to its initial value $t_{lt}(0) \simeq 2 \tacc (1+f_{M,0})$ \revmnras{(see Eq. \ref{eq:tlt-instant})}. b) Trajectory of the disc in the $M_{D}-\dot{M}_{*}$ plane, in the constant $\alpha$ case, and in the $\Sigma_{\text{c}}$-dependent wind torque case. The time evolution is pictured by the markers which represent the discs at $t/\tacc=0, 0.5,1,2,4,8,16$, and $32$. The dashed lines in the $\Sigma_{\text{c}}$-dependent wind torque case picture the trajectory of the disc during its fast dispersal, as defined as $t > 0.9 t_{dis}$. The wind parameters are set to $\lambda=3$ and $r_c(0)/r_{in}=50$.}
    \label{fig:MD-Macc-traj} 
\end{figure*}

The evolution of an individual disc in the $\MaccMd$ plane can be studied using the disc lifetime\footnote{Also called 'disc age' by \citet{2012MNRAS.419..925J}. Here we follow the nomenclature of \citet{2017MNRAS.472.4700L} as in the wind-driven case this quantity is not necessarily related to the age of the disc, as shown in Fig. \ref{fig:MD-Macc-traj}-a.}
\begin{equation}
    t_{lt}(t) \equiv \frac{M_D(t)}{\dot{M_{*}}(t)}.
    \label{eq:tlt-def}
\end{equation}
The main advantage of this quantity is to be independent on the initial disc mass and to be directly comparable with observations. For both classes of self-similar solutions the disc lifetime is
\begin{equation}
\begin{split}
        t_{lt}(t)& \revmnras{= 2 \frac{1+\psi}{1+\psi+2\xi} t_{\text{acc}}(t) (1+f_{M}(t))}\\
       & \revmnras{\simeq 2 t_{\text{acc}}(t) (1+f_{M}(t))~~\textnormal{for}~~\xi \ll 1 ~~\textnormal{or}~~ \psi \gg 1},
\end{split}
    \label{eq:tlt-instant}
\end{equation}
where $t_{\text{acc}}(t)$ is the instantaneous accretion timescale defined as
\begin{equation}
t_{\text{acc}}(t) \equiv \frac{r_c(t)}{3 \epsilon_c c_{s,c} \tilde{\alpha}},
\label{eq:tacc-instant-def}
\end{equation}
where $\epsilon_c$ is the disc aspect ratio and $c_{s,c}$ is the sound speed evaluated at $r = r_c(t)$. This is a generalisation of the initial accretion timescale $\tacc = t_{acc}(t=0)$. In particular, the instantaneous accretion timescale varies with time due to the evolution of $r_{c}(t)$ and $\tilde{\alpha}(t)$ as
\begin{equation}
   t_{acc}(t)= \frac{r_c(t)}{r_c(0)} \left(\frac{\tilde{\alpha}(t)}{\tilde{\alpha}(0)} \right)^{-1} \tacc~~,
   \label{eq:tacct-bis}
\end{equation}
where we recall that $\tilde{\alpha}= \alphaDW+\alphaSS$.


Figure \ref{fig:MD-Macc-traj}-a shows that the time evolution of the disc lifetime $t_{lt}(t)$ depends critically on the accretion mechanism. In the fiducial solution ($\psi=\infty$, $\omega=0$), $f_M(t)$, $\alphaDW(t)$, and $r_c(t)$ are constant, resulting in a constant disc lifetime
\begin{equation}
    t_{lt}(t) = 2 \tacc (1+f_{M,0}).
    \label{eq:disc-lifetime-fiducial}
\end{equation}
In contrast, in the general case, the disc lifetime $t_{lt}(t)$ varies with time. This is ultimately due to a variation of $r_c(t)$ in the hybrid solutions, and of $\alphaDW(t)$ in the $\Sigma_{\text{c}}$-dependent wind torque solutions. 

For the hybrid solutions, disc spreading results in an increase in $t_{lt}(t)$ driven by the increase in both $\tacct$ and, to a lesser extent, $f_{M}(t)$. Quantitatively, the disc lifetime grows over a viscous timescale $t_{\nu,0} = (1+\psi) \tacc$ as
\begin{equation}
\begin{split}
       t_{lt}(t)&  = 2\tacc (1+f_{M,0}) \frac{\psi+1}{\psi+1+2\xi} \left( 1+\frac{t}{(1+\psi) \tacc} \right)^{1+\xi} \\
       & \simeq  2\tacc (1+f_{M,0}) \left( 1+\frac{t}{(1+\psi) \tacc} \right)~~\textnormal{for}~~\xi \ll 1 \revmnras{~~\textnormal{or}~~ \psi \gg 1}.
\end{split}
\end{equation}
In the viscous regime ($t\gg t_{\nu,0}$) and for $\xi \ll 1$, the disc lifetime is proportional to the disc age $t_{lt} \simeq  2 t (1+f_{M,0})/(1+\psi)$. This is a generalization of the results found for viscous discs \citep{2012MNRAS.419..925J,2017MNRAS.468.1631R}.


In the $\Sigma_{\text{c}}$-dependent wind torque solutions, $\alphaDW$ increases with time, resulting in a decrease in $\tacct$ (see Eq. (\ref{eq:tacct-bis})) and so in $t_{lt}(t)$. Quantitatively, the disc lifetime declines linearly with time as
\begin{equation}
    t_{lt}(t) = 2\tacc (1+f_{M,0}) \left( 1-\frac{\omega}{2 \tacc} t \right)
\end{equation}
before full dispersal.

\subsubsection{Evolutionary track}
\label{subsec:evo-tracks}

The evolutionary track of a disc in the $\MaccMd$ plane is shown in Fig. \ref{fig:MD-Macc-traj}-b. In both classes of solution, the accretion rate is a power-law  of the disc mass (i.e. straight tracks in Fig. \ref{fig:MD-Macc-traj}-b). In the fiducial case, $\dot{M}_{*}$ is proportional to $M_D$ as the disc lifetime is constant along the evolutionary track. On either side of this reference track, the hybrid solutions follow tracks that are steeper as the disc lifetime increases with time, whereas the $\Sigma_{\text{c}}$-dependent wind torque solutions describe tracks that are flatter, as the disc lifetime decreases with time.

In the hybrid solutions the evolutionary track is given by
\begin{equation}
    \dot{M}_* = \dot{M}_{*,0} \left(\frac{M_D}{M_0}\right)^{\frac{\psi+3+4\xi}{\psi+1+2\xi}},
\end{equation}
where $\dot{M}_{*,0}$ is the initial accretion rate (see Eq. (\ref{eq:Mdot-initi})). Therefore, the power-law index decreases, from the pure viscous case ($\dot{M}_{*} \propto M_D^{3}$) to the fiducial pure wind case ($\dot{M}_{*} \propto M_D$) as shown in Fig. \ref{fig:MD-Macc-traj}-b. In the viscous case, the steeper drop in $\dot{M}_*$ along the track it ultimately due to the disc spreading. We note that allowing $\alphaSS$ and $\alphaDW$ to vary as a power-law, we recover the degeneracy between $\gamma$ and $\psi$. For example, classical turbulent solutions with $\alphaSS \propto r^{-1+\gamma}$ and $T \propto r^{-1/2}$ gives $\dot{M}_* \propto M_D^{5-2\gamma}$ meaning that evolutionary tracks are shallower for larger values of $\gamma$ \citep{2017MNRAS.472.4700L}.

For the $\Sigma_{\text{c}}$-dependent wind torque solutions
\begin{equation}
   \dot{M}_* = \dot{M}_{*,0} \left(\frac{M_D}{M_0}\right)^{1-\omega}.
\end{equation}
Trajectories are flatter than $\dot{M}_* \propto M_D$, with a power law index ranging from $1$ in the fiduical case ($\omega=0$) to 0 ($\omega=1$). 

As time increases, a disc runs along its own evolutionary track. The markers in Fig. \ref{fig:MD-Macc-traj}-b highlight the location of the disc at different time steps. For the same accretion timescale $\tacc$, discs evolve faster for flatter tracks as the decline of accretion rate is shallower. In particular, for hybrid solutions, the evolution of the disc along the track is faster for stronger wind torque (i.e., higher value of $\psi$). In the $\Sigma_{\text{c}}$-dependent wind torque solutions, the disc are dispersed at a finite time as the drop in accretion rate is slower than that of the disc mass (tracks flatter than $\dot{M}_*\propto M_D$). In this case, a disc spends most of its lifetime in the high mass part of its track and then runs rapidly through the low mass part during a short dispersal phase. This is better shown in Fig. \ref{fig:MD-Macc-traj}-b, where the dotted line corresponds to the track swept in the last $1\%$ of the effective lifetime of the disc.

\subsubsection{Isochrones}
\label{subsec:isochrones}
\begin{figure}
	\includegraphics[width=\columnwidth]{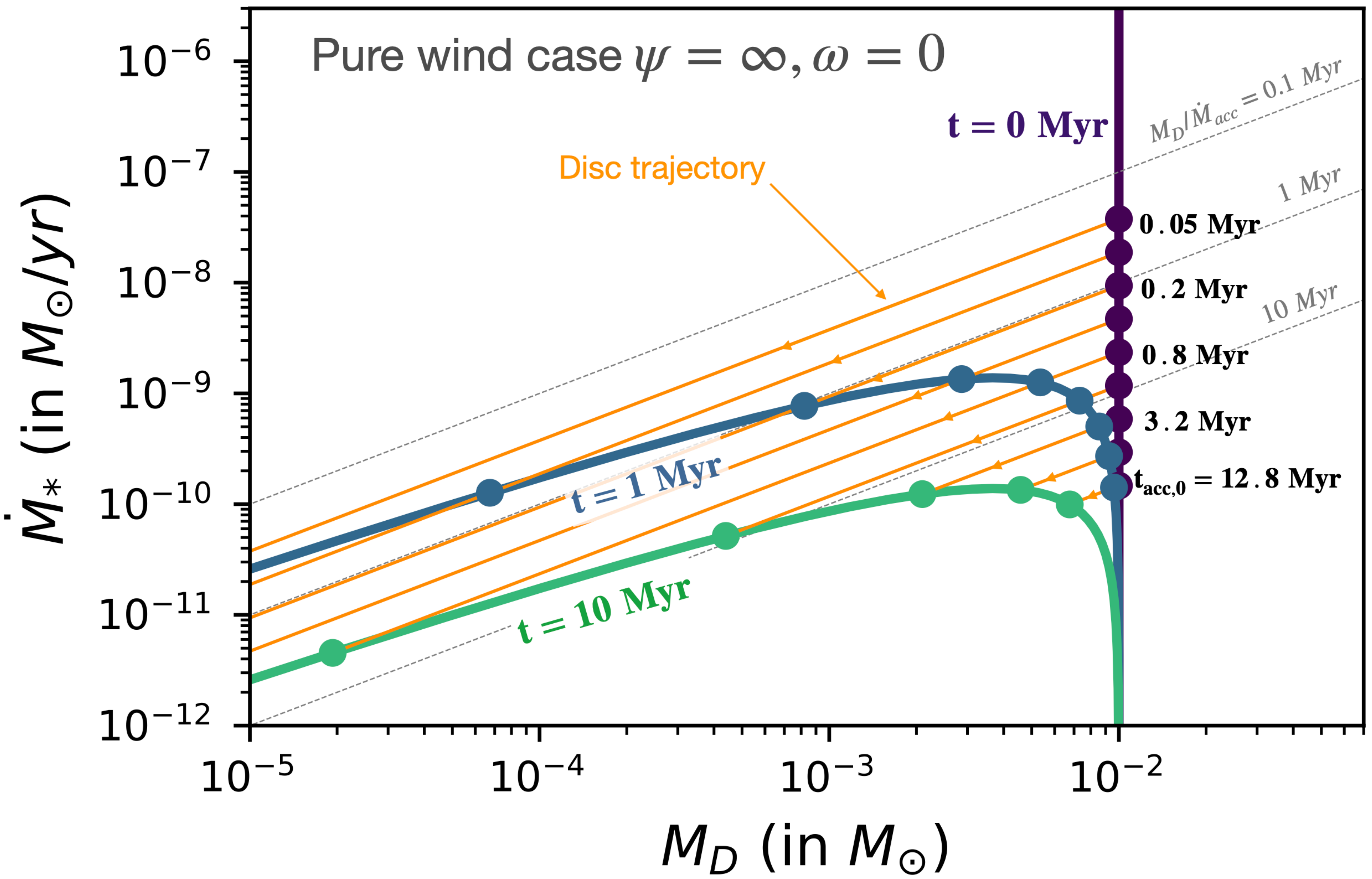}
    \caption{Isochrones and evolutionary tracks in our fiducial case ($\psi=+\infty, \omega=0$) for an initial disc mass of $M_0 = 10^{-2} M_{\odot}$. The orange lines are the evolutionary tracks of a sample of disc starting with different values of $\tacc$ as indicated along the $t=0$ isochrone. \revmnras{The wind parameters are set to $\lambda=3$ and $r_c/r_{in} = 50$, which corresponds to an ejection-to-accretion ratio of $f_{M,0}=1.7$.}}
    \label{fig:isochrones-fiducial} 
\end{figure}
\begin{figure*}
	\includegraphics[width=1.95\columnwidth]{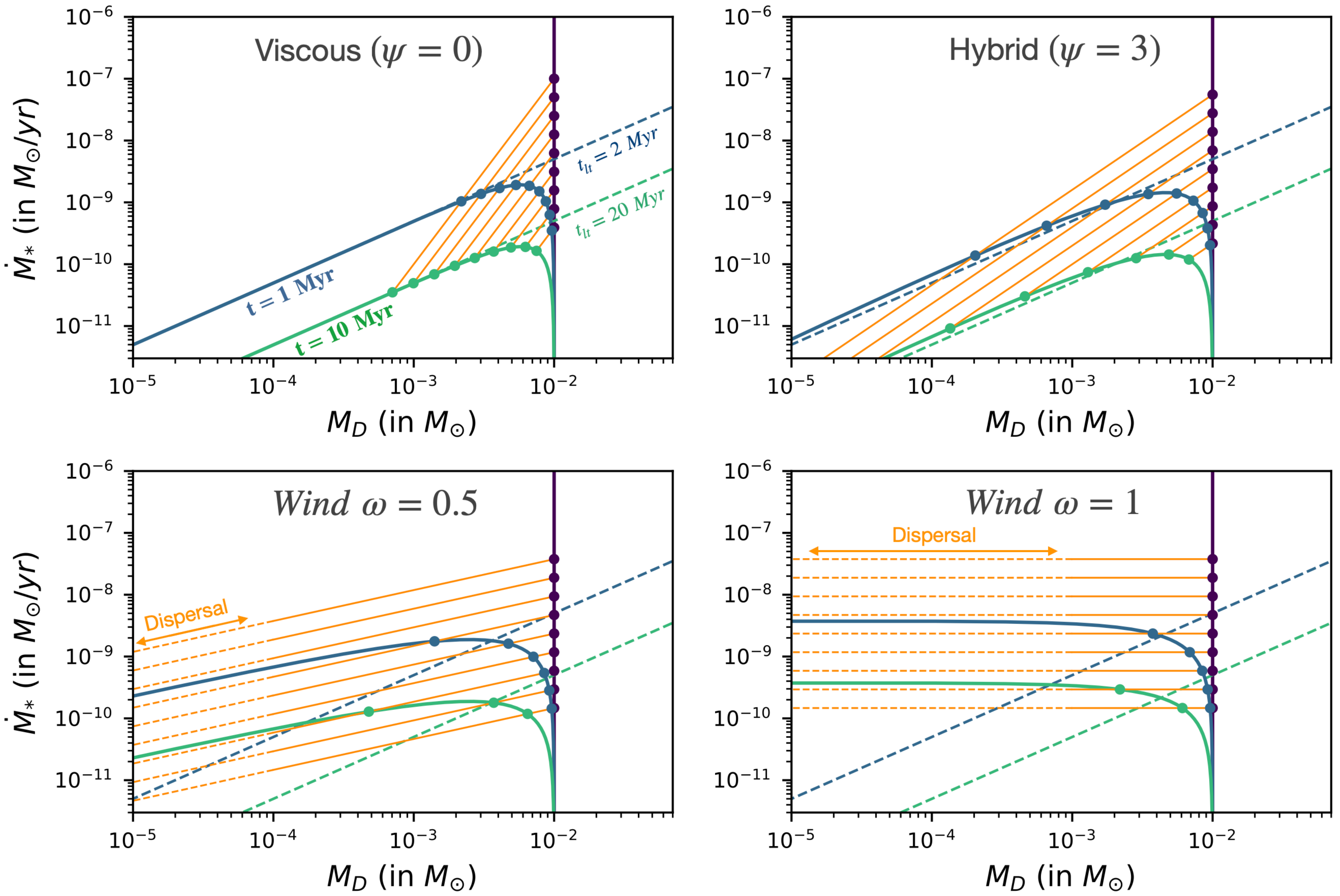}
    \caption{Isochrones and evolutionary tracks depending on the parameters of the solutions for an initial disc mass of $M_0 = 10^{-2} M_{\odot}$. The isochrones are plotted in solid lines with the age $t$ that is color coded ($t=0, 1, 10$~Myr). By definition, all the discs located along an isochrone start with the same disc mass but a different initial $\tacc$ such that their initial accretion rate is different. \revmnras{Discs of different $\tacc$ are highlighted by markers with $\tacc$ ranging from $0.05$ to $12.8$ Myr as indicated in Fig. \ref{fig:isochrones-fiducial}.} \revmnras{The wind parameters are set to $\lambda=3$ and $r_c/r_{in} = 50$, which corresponds to a mass ejection-to-accretion ratio of $f_{M,0}=1.7$ in the pure wind cases (bottom panels) and $f_{M,0}=1$ for $\psi=3$ (top right panel)}. The effect of the mass loss rate is to lower the accretion rate by a factor $(1+f_{M,0})$. Therefore, a change in the relative mass loss rate would simply shift the isochrones vertically.}
    \label{fig:isochrones-grid} 
\end{figure*}

When studying a population of discs of a similar age, the concept of isochrone is particularly valuable. An isochrone describes the locus in the $M_D-\Macc$ plane of a sample of discs that share the same initial mass $M_0$ but have different initial accretion timescales $\tacc$. The shape of an isochrone is the result of the evolution of each disc of different $\tacc$ along its own evolutionary track. By construction, it does not depend on $\tacc$ but on the age of the disc sample $t$, and on the initial disc mass $M_0$.

Figure \ref{fig:isochrones-fiducial} shows the isochrones for the fiducial solution. The high mass part of the isochrone is vertical. This corresponds to discs that have long accretion timescales ($\tacc \gg t$) and thus discs that have not evolved yet and have a low initial accretion rate ($\dot{M}_{*}(t=0) \propto 1/\tacc$). At lower disc mass, the isochrone bends toward the x-axis. In this region, discs have evolved ($\tacc \lesssim t$). As discussed above (Sec. \ref{subsec:disc-lifetime}), individual discs have a constant disc lifetime $t_{lt}$ that is proportional to $\tacc$ (see Eq. (\ref{eq:disc-lifetime-fiducial})). Therefore, discs that are in the low mass regime have also a small disc lifetimes (see grey lines in Fig. \ref{fig:isochrones-fiducial}). Quantitatively, by eliminating $\tacc$ in Eq. (\ref{eq:disc-mass-fiducial}) and (\ref{eq:acc-rate-fiducial}), the shape of the isochrone in the fiducial case yields
\begin{equation}
    \dot{M}_* = \frac{1}{(1+f_{M,0}) t} M_D \ln{M_0/M_D}.
\end{equation}
Interestingly, the isochrone depends on the mass ejection-to-accretion ratio $f_{M,0}$ as $\dot{M}_*$ is lowered by the wind mass loss rate. The effect of wind mass-loss rate is simply to shift the disc isochrones downward. This means that $f_{M,0}$ could be constrained from disc demographics.

Figure \ref{fig:isochrones-grid} shows that the shape of the isochrones depends critically on the transport of angular momentum. This difference is the result of different slopes of the disc evolutionary tracks (see orange lines, Fig. \ref{fig:isochrones-grid}). In the pure viscous case ($\psi=0$), we recover the result of \citet{2017MNRAS.472.4700L} that the isochrone converges at low disc masses towards a linear relation between $\dot{M}_*$ and $M_D$. Quantitatively, for $t \gg \tacc=t_{\nu,0}$ and a spatially constant $\alphaSS$, the disc lifetime does not depend on the initial accretion timescale but on the disc age as $t_{lt} \simeq 2 t$, which is by definition the same along the isochrones. Interestingly, this asymptotic behaviour does not depend on the power-law index of $\alpha$. In fact, for $t \gg t_{\nu,0}$ and $\alphaSS \propto r^{-1+\gamma}$, the disc lifetime is $t_{lt} \simeq 2 (2-\gamma) t$ \citep{2012MNRAS.419..925J,2017MNRAS.468.1631R}. In other words, in the turbulent case, the isochrones align along a constant $M_D/\dot{M}_*$ line that is about the age of the isochrones.

In the hybrid case, the isochrones share features of both pure wind and pure viscous case. 
Quantitatively, the isochrones write:
\begin{equation}
\begin{split}
       \dot{M}_* &= \frac{\psi+1+2\xi}{2(1+f_{M,0}) t} M_D \left[ 1- \left(\frac{M_D}{M_0}\right)^{\frac{2}{\psi+1+2\xi}}\right]  \left(\frac{M_D}{M_0}\right)^{\frac{2\xi}{\psi+1+2\xi}} \\
       & \simeq \frac{\psi+1}{2(1+f_{M,0}) t} M_D \left[ 1- \left(\frac{M_D}{M_0}\right)^{\frac{2}{\psi+1}}\right]~~\textnormal{for}~~\xi \ll 1.
\end{split}
\end{equation}
The isochrone can be decomposed in two parts. In the high mass part, corresponding to discs that are in an early stage of evolution, the isochrone is similar to that of the pure wind case as viscous spreading does not impact the disc evolution. In the low mass part, the isochrone aligns along a constant $M_D/\dot{M}_*$ line as viscous spreading takes over. This defines the viscous regime for which $t_{lt} \simeq 2 t (1+f_{M0})/(1+\psi)$, regardless of the initial $\tacc$.

In the $\Sigma_{\text{c}}$-dependent wind torque solutions, the bending of the isochrone is less pronounced as disc evolutionary tracks are flatter. Quantitatively, the isochrones writes
\begin{equation}
    \dot{M}_* = \frac{1}{\omega (1+f_{M,0}) t} M_D \left(\left(\frac{M_D}{M_0}\right)^{-\omega}-1\right).
\end{equation}
The lower mass part of the isochrone corresponds to discs that are in the process of being dispersed $t_{disp}  \simeq t$. This implies that at low disc mass, the isochrones converges towards the evolutionary track of the disc that is being dispersed and follows $\dot{M}_* \propto M_D^{1-\omega}$. However, the probability to find a disc that is in the low mass part of the isochrone is unlikely since it corresponds to a short lived state. The isochrone should then be truncated at low $M_D$. 




\section{Discussion}
\label{sec:disscu}

\subsection{Observational constraints of MHD disc winds}
\label{subsec:obs-lambda}

The direct observation of MHD disc wind candidates provides us with first constraints on wind parameters.
The measurement of the rotation signature, the axial velocity, and the mass-loss rate of outflows allows one to determine the launching region of the MHD disc wind, the mass ejection-to-accretion ratio, the magnetic lever arm parameter, and the fraction of angular momentum extracted vertically \citep{2003ApJ...590L.107A,2006A&A...453..785F,2020A&A...640A..82T}. 

ALMA has already unveiled rotation signature in outflows from embedded protostars, for which the mass loss and accretion rates are the highest \citep{2016Natur.540..406B,2017A&A...607L...6T,2020A&A...634L..12D}. Magnetic lever arm parameters are consistently found to be low, typically between $\lambda=1.6-5$. Mass-loss rates are about the accretion rates ($f_{M} \simeq 1$) and, at least in the HH212 protostar, compatible with an MHD disc wind that extracts most of the angular momentum required to drive accretion from the bulk part of the disc \citep[out to $40$~au,][]{2020A&A...640A..82T}.

Regarding Class II discs, early studies have unveiled rotation signatures in atomic jets ($\gtrsim 50~$km/s) suggestive of MHD disc winds launched from the inner regions of the discs ($\lesssim$ few $au$) with magnetic lever arm of $\lambda \gtrsim 4$ and low mass loss rates of $\dot{M}_W \simeq 0.1 \dot{M}_{*}$ \citep[see][and reference therein]{2006A&A...453..785F}. However, the presence of MHD disc winds launched from the bulk part of the disc remains largely unconstrained. 
\citet{2018A&A...618A.120L} found a massive rotating outflow emanating from the HH30 T Tauri star suggestive of an MHD disc wind that has a mass loss rate about the accretion rate, a low magnetic lever arm ($\lambda \simeq 1.6$), and a launching radius about a few au. However, the appearance of the outflow as a V-shape cavity suggests that it is not pristine wind material but rather shocked gas tracing the interaction of the wind with the jet \citep{2018A&A...614A.119T}, also seen in DG Tau B and HH212 \citep{2020A&A...634L..12D,2021ApJ...907L..41L}. This perturbation of the wind by a fast jet might bias the determination of the angular momentum extracted by the wind or the launching region.
Overall, it remains to be determined if the scarcity of clear CO winds launched from extended regions of Class II discs is due to the absence of MHD disc winds, the low sensitivities reached by current observations (in particular in $^{12}$CO lines), or the destruction of CO in the wind. In fact observations conducted in the visible at high spectral resolution provide further evidence for a slow atomic component ("narrow low velocity components") that could either trace photoevaporative winds or MHD disc winds \citep{2016ApJ...831..169S,2020ApJ...903...78P}. Still the lack of spatial and spectral resolution prevent us from deriving robust estimates of wind mass loss rate and $\lambda$. Detailed modeling including chemistry and thermal balance is warranted to further interpret these data. 

To conclude, observations of young protostars shows evidence of MHD disc winds with $\lambda \simeq 1.6-5$ and $f_{M} \simeq 1$. MHD winds launched from the bulk part of Class II discs are largely unconstrained. Further observational campaign at high spectral resolution, combined with details astrochemical modelling are required to further test the presence of MHD disc wind and constrain the wind parameters. In the absence of such constrains, disc demographics can turn out to be a complementary and independent approach to test the presence of MHD disc winds.


\subsection{Observational perspectives on disc demographics}
\label{sec:obs-perspective}

The analytical solutions presented in this work pave the road to assess the role of MHD disc winds from the observations of disc populations. In this section, we discuss the analysis that can be conducted to confront this simple analytical solutions to observations. This also allows us to give first quantitative estimates of the key parameters of the solutions presented in this work. A first detailed comparison between the $\Sigmac$-dependent wind torque solutions and the observations is presented in a companion paper (Tabone et al., subm.).

\subsubsection{Correlating source properties with source age}

Our secular model predicts the evolution of disc radius, mass, and accretion rate with time. A possible approach to test our disc evolution models would then be to directly study the evolution of source properties with the source age. This approach has already been adopted to analyse the decline of the accretion rate with source age in the viscous paradigm \citep{1998ApJ...495..385H,2012A&A...538A..64C}. However, the simplicity of this approach hides important caveats. First of all, the unknown systematic uncertainties on estimation of stellar age leads to systematic errors on the dependency of source parameters with source age \citep{2014prpl.conf..219S}. Secondly, this approach does not consider disc dispersal that could drive the observed evolution of the disc properties by preferentially removing discs with specific properties \citep[see e.g.][for the viscous case]{2020MNRAS.492.1120S}. This "survivorship bias" is expected to play a role for source ages larger than the disc dispersal time ($\simeq 2-3~$Myr). In fact, in a companion paper, we show that for the $\Sigmac$-dependent wind torque solutions, the median disc mass declines due to the dispersal of the disc with the highest accretion rates whereas the accretion rate of individual disc is constant over time for $\omega=1$ (see Supplementary material in Tabone et al., subm.).

\rev{In order to mitigate the first caveat, it is now common to study the accretion rate and disc properties (mass, size...) of a given cluster, assuming a similar age for the disc population \citep[e.g.,][]{2016A&A...591L...3M,2017ApJ...847...31M}. The second caveat highlights the need for a disc population synthesis approach. Whereas the present work provides models required to build such synthetic populations, this approach is beyond the scope of this work.}

\subsubsection{Disc size}

The simplest distinctive feature between wind-driven and turbulence-driven evolution lies in the evolution of disc size. In particular, our secular model including turbulent transport predicts that the disc characteristic radius increases with time on a time-scale of $t_{\nu,0}=\tacc (1+\psi)$. In the case of $\Sigma_{\text{c}}$-dependent wind torque solutions, turbulence is neglected. However, one can expect that in the presence of a non-vanishing level of turbulence, the disc spreading would also happen on a viscous timescale $t_{\nu,0}$. 


Therefore, the measurement of disc sizes could constitute a discriminant test between wind-driven and turbulence driven accretion and provide valuable constraints on the viscous timescale. However, the measurement of the disc characteristic radius remains challenging. Continuum sizes have been measured toward a large sample of sources in star forming regions of different ages. However, the characteristic size of dust emission is mostly indicative of dust growth and drift processes rather than the true radial profile of the gas \citep{2019MNRAS.486L..63R,2019MNRAS.486.4829R,2019ApJ...878..116P}. As such, the observation of a rather constant continuum size between Class 0, I and II discs of about $\simeq 50~$au \citep{2020ApJ...890..130T,2021A&A...649A..19S} should be interpreted with caution.

Alternatively, characteristic emission radii of CO rotational lines have been measured, though on a smaller sample of sources due to sensitivity issues \citep{2017ApJ...851...85B,2018ApJ...859...21A,2021A&A...649A..19S}. However, this characteristic radius is indicative of the region of the disc where the CO lines become optically thin, either due to a drop in the surface density profile or a drop in CO abundance due to e.g., photodissociation \citep{2019A&A...629A..79T}. Nevertheless, viscous models coupled to detailed modelling confirms that in general the observed radius, although in general not a good tracer of true characteristic radius, is expected to expand with time with a rate increasing with the magnitude of the viscosity \citep{2020A&A...640A...5T}. Therefore, detailed modelling including chemistry, thermal balance, and radiative transfer are required to derive the true characteristic radius $r_c$ from the observed CO emission profile. Moreover, deep ALMA observations in the gas are still missing to measure disc gas size in the majority of the discs.


\subsubsection{Disc lifetime and initial accretion timescale}
\label{subsec:disc-lifetime-tacc}

\begin{figure*}
\includegraphics[width=2\columnwidth]{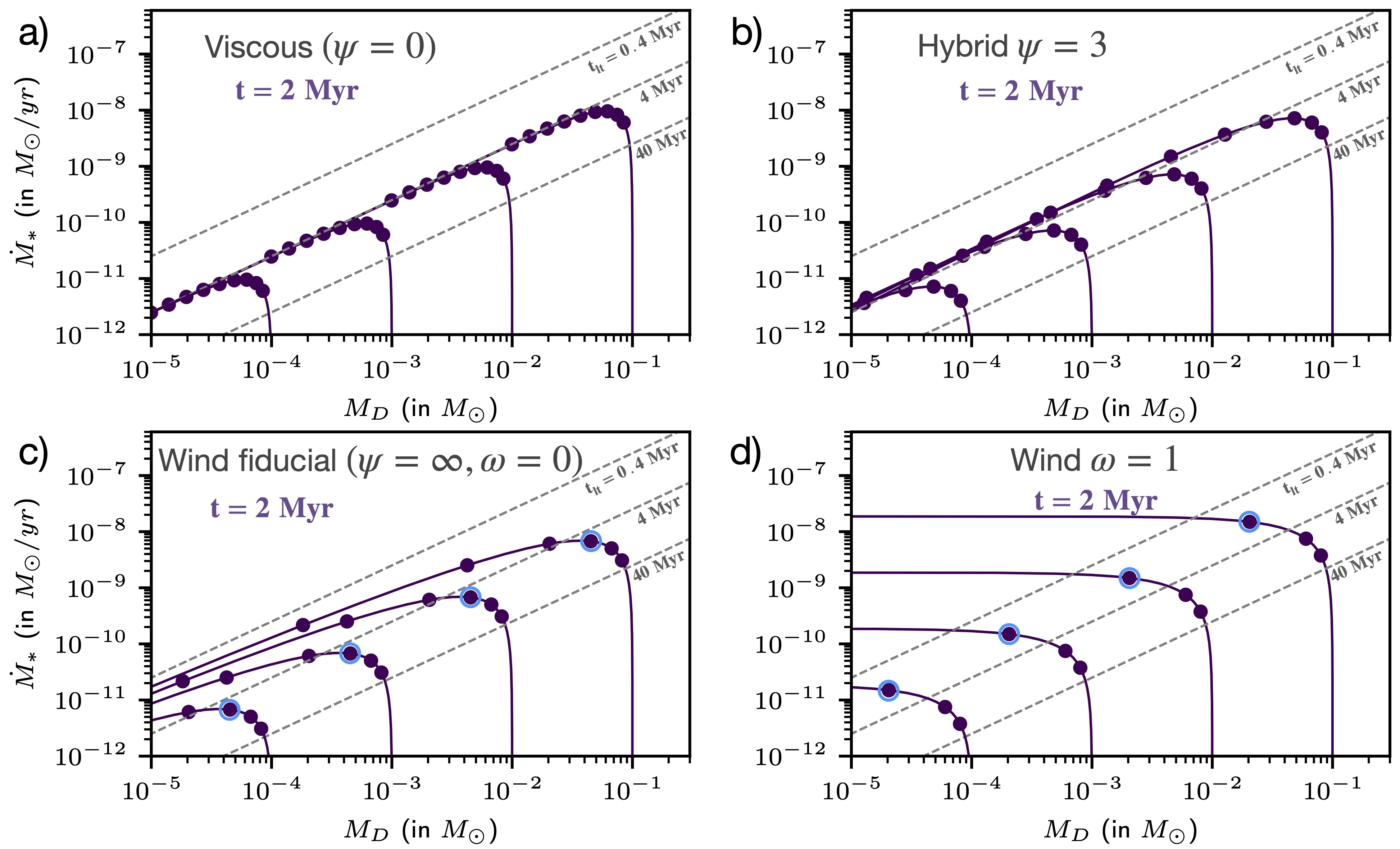}
    \caption{Distribution of the discs in the $\MaccMd$ plane after $t=2~$Myr depending of the solutions and starting from a distribution in $M_0$ and $\tacc$. The isochrones are plotted in each panel in solid lines for different initial disc mass ($M_0=10^{-4}, 10^{-3},10^{-2},10^{-1} M_{\odot}$). The dots represent individual discs that have started with $\tacc$ ranging from 0.02 to $5$ Myr with logarithmic steps. The cyan circles in the bottom panels (pure wind cases) highlight discs born with $\tacc = 2.5~$Myr. In the $\Sigmac$-dependent wind torque solution (panel d), discs born with $\tacc < t_{\text{disp}}$ have dispersed.}
    \label{fig:isochrones-grid-pop} 
\end{figure*}

As shown in Sec. \ref{sec:MD-Macc}, the observed disc lifetime $t_{lt}=M_D/\dot{M}_*$ gives an upper limit to the instantaneous accretion timescale $\tacct$ (see Eq. (\ref{eq:tlt-instant})). If $f_M \lesssim 1$, $t_{lt}(t) \simeq 2 \tacct$. ALMA and XShooter surveys found $t_{lt}\simeq 3~$Myr in the 2Myr old clusters of Lupus and Chamaeleon implying an instantaneous accretion timescale of about $t_{\text{acc}}(2~Myr) \lesssim 1.5~$Myr. However, this value is not necessarily the initial accretion timescale $\tacc$. In the hybrid case, $\tacc \lesssim 1.5~$Myr as $\tacct$ increases with time. In the pure wind case, $\tacc \gtrsim 1.5~$Myr as $\tacct$ decreases with time (see Fig. \ref{fig:MD-Macc-traj}-a). \rev{Still, as long as the wind torque dominates, the instantaneous accretion timescale gives a rough estimate of $\tacc$. This is because even in the $\Sigmac$-dependent torque solutions, $\tacct$ is almost constant during most of the disc lifetime and decreases substantially only when the disc is on the verge of dispersal. We can therefore anticipate that typical values of the initial accretion timescale $\tacc$ are of the order of a Myr. This contrasts with viscous accretion for which $\tacc$ can be much shorter than $t_{lt}(t)$ as $t_{lt}(t) \simeq t$ for $t \gg \tacc =t_{\nu,0}$.}

Provided an independent measurement of $r_c(t)$ is available, one could then derive an estimate of $\tilde{\alpha}$. Quantitatively, for $\epsilon(r) = 0.1 (r/50\rm{au})^{1/4}$,
\begin{equation}
\tilde{\alpha}(t) = 1.8 \times 10^{-3} \left( \frac{\tacct}{10^6 yr}\right)^{-1} \left(\frac{r_c(t)}{50~au}\right) \revmnras{\left(\frac{M_*}{1M_{\odot}}\right)^{-1/2}}.
\label{eq:alpha-numerical-value}
\end{equation}
Assuming a typical disc characteristic radius of $r_c = 50$~au, we then derive $\tilde{\alpha}(t=2Myr) \simeq 10^{-3}$. Again, if accretion is driven by a wind, this value provides us with a rough estimate of the initial value of $\alphaDW$. However, these constraints do not allow one to distinguish between to two accretion mechanisms.



\subsubsection{$\MaccMd$ correlation}

Combining ALMA and VLT/XShooter surveys, \citet{2016A&A...585A.136M} and \citet{2017ApJ...847...31M} found the first evidence for a correlation between accretion rate and disc mass in Lupus and Chamaeleon with a nearly linear relationship and a large scatter of about $1~$dex around this trend. 

\rev{In this section, we discuss under what conditions our solutions predict such features in the $\dot{M}_*-M_D$ plane. A detailed investigation of this question requires to build realistic disc population synthesis models. Here, we use a toy population model, starting from a sample of disc with four initial mass ranging from $M_0=10^{-4}$ to $10^{-1} M_{\odot}$, and 9 values of $\tacc$ ranging from $0.02$ to $5$~Myr. These values are in line with the masses of Class I discs \citep{2020A&A...640A..19T,2020ApJ...890..130T} and the constraints on $\tacc$ obtained in Sec. \ref{subsec:disc-lifetime-tacc}. A realistic disc population model based on the $\Sigma_{\text{c}}$-dependent wind torque solutions are presented in a companion paper (Tabone et al., subm.).}

\rev{Figure \ref{fig:isochrones-grid-pop} shows the location of the synthetic discs in the $\MaccMd$ plane after $2$~Myr for different solutions. We recall that discs that have started with the same masses are located along the same disc isochrone (purple curves). In the viscous case (Fig. \ref{fig:isochrones-grid-pop}-a), we recover that the majority of the discs align along the same line of constant $t_{lt} \simeq 4~$Myr, regardless of the initial disc mass. This is a typical feature already discussed by \citet{2017MNRAS.472.4700L,2017MNRAS.468.1631R}. Because of the viscous spreading, that takes over disc evolution for $t \gtrsim \tacc$, the initial condition of the disc are "reset" as the disc lifetime converges toward $t_{lt} \simeq 2 t$, regardless of the initial disc mass or $\tacc$. The corollary is that the dispersion of the data around the main trend shrinks for discs with $t \gtrsim \tacc= t_{\nu,0}$. }

\rev{In the hybrid case (Fig. \ref{fig:isochrones-grid-pop}-b), we still recover a tight correlation between $\dot{M}_*$ and $M_D$ for discs that are in the viscous regime ($t \gtrsim t_{\nu,0} = (1+\psi)\tacc$), regardless of the initial disc mass or $\tacc$. Because the viscous timescale is longer as $t_{\nu,0} = \tacc (1+\psi)$, a slightly larger fraction of disc are off the main trend. On average, the disc masses have also declined more as wind accretion is more efficient than viscous accretion.}

\rev{Figures \ref{fig:isochrones-grid-pop}-c-d show that in the case of pure wind-driven accretion, discs starting with different $M_0$ and $\tacc$ do not align along the same constant $M_D/\dot{M}_*$ line after 2~Myr. In other words, the isochrones do not overlap in the low mass part, in contrast to the viscous case. This essential feature is due to the absence of disc spreading: each disc keeps the memory of its initial conditions. It follows that in contrast to the viscous cases, a correlation between $\dot{M}_*$ and $M_D$ is not expected for arbitrary initial conditions.}

\rev{However, a correlation between $\dot{M}_*$ and $M_D$ can be obtained for specific initial conditions. 
In particular, starting from a large spread in $M_0$ but a single value of $\tacc$, we predict a perfectly linear correlation between $M_D$ and $\dot{M}_*$ (see points highlighted in Figs. \ref{fig:isochrones-grid-pop}-c-d). The initial distribution of $M_0$ is also crucial for the $\MaccMd$ correlation. In the extreme case of a single initial disc mass, the disc population is confined to an isochrone, which is not a linear function. For  the initial distributions of $\tacc$ and $M_0$ adopted in Figs. \ref{fig:isochrones-grid-pop}-c-d we recover a correlation between $\dot{M}_*$ and $M_D$. As a rule of thumb, a correlation with a nearly linear relationship is obtained if the distribution of $\tacc$ is the same for all the values of the initial disc mass $M_0$ and with a rather broad distribution of initial disc mass. Further investigation should determine how much correlation between the initial $\tacc$ and $M_0$ can be introduced before altering significantly the relationship. This illustrates the fact that for wind-driven accretion, the distribution of the discs in the $M_D-\dot{M}_*$ plane carries information about the initial conditions of the disc population set during the protostellar phase.}  

\rev{Another striking difference between viscous and wind accretion in Fig. \ref{fig:isochrones-grid-pop} is the large dispersion of the discs around the mean trend in the latter case. Indeed, for wind-driven accretion, discs keep the memory of their initial conditions: a dispersion in the initial disc lifetime $t_{lt}= M_D/\dot{M}_*$, which reflect a distribution in $\tacc$, is maintained during the evolution of the population. For the fiducial solution, the value of $t_{lt}$ of each of the disc is constant over time and proportional to $\tacc$ (see Sec. \ref{subsec:disc-lifetime} and Fig. \ref{fig:MD-Macc-traj}-a). As a result, a large dispersion in $\tacc$ results in a dispersion in $t_{lt}$ at any time. For the $\Sigma_{\text{c}}$-dependent wind torque solution, the situation is more complicated as $t_{lt}$ deceases with time whereas discs born with the lowest $t_{lt}$ (i.e. lowest value of $\tacc$) disperse first. Detailed study using disc population synthesis are required to better determined how the dispersion in the observed disc lifetime $t_{lt}$ reflect the initial distribution of $\tacc$.}

In other words, in the wind case, the $\MaccMd$ correlation reflects the initial properties of the disc population. If $M_0$ and $\tacc$ have a large spread and if the distribution of $\tacc$ is independent on $M_0$, a linear correlation with a large dispersion around the mean trend is predicted. 



\subsubsection{Disc dispersal}
\label{subsubsec:disc-dispersal}

Disc dispersal is one of the most fundamental feature of disc evolution. Extensive surveys in the optical and infrared have shown that discs disperse after a typical time of about $t_{disp} = 2-3~$Myr in a short dispersal timescale of $\Delta t_{disp} \lesssim 0.5$~Myr \citep{1995ApJ...450..824S,1996AJ....111.2066W}.


\citet{2013ApJ...778L..14A} have shown that MHD wind accretion can induce fast disc dispersal if the magnetization increases with time, due to a slower dissipation of the magnetic field compared to the gas \citep[see also][]{2016ApJ...821...80B,2016A&A...596A..74S}. In contrast with the viscous scenario, wind-driven accretion would then account for both disc accretion and disc dispersal. Our $\Sigma_{\text{c}}$-dependent wind torque solutions provide us with a simple framework to describe this process and analyse the observational data. In particular, we predict that the dispersal time $t_{disp}$ depends on $\tacc$ and $\omega$ as
\begin{equation}
    \tdisp = \frac{ 2\tacc}{\omega}.
\end{equation}

Because the disc dispersal is connected to the accretion process, the constraints on $\tacc$ obtained from disc dispersal prescribes the accretion properties. One can already notice that for $\omega=1$, the typical dispersal time of $\tau \simeq 2.5~$Myr implies a typical accretion timescale of $\tacc \simeq 1~$Myr. This is in line with the value of $\tacc$ estimated from the observed median disc lifetime of $t_{lt} \simeq 3~$Myr (see Sec. \ref{subsec:disc-lifetime-tacc}). However, disc population synthesis models are required to properly reproduce both disc dispersal and the accretion properties. In a companion paper, we follow this argument and show that disc dispersal and accretion properties can be simultaneously reproduced by the $\Sigma_{\text{c}}$-dependent wind torque solutions.

The solutions of constant $\alphaDW$ might seem to be incompatible with disc dispersal as disc mass and accretion rate never drop to zero. However, the drop in $M_D(t)$ and $\dot{M}_*(t)$ is steep and the disc will be undetectable in the IR and in the visible (accretion signature) after a finite time, which depends on the detection limits and on the exact values of the disc parameters ($M_0$, $\tacc$, $\psi$). Detailed models are thus required to assess if solutions of constant $\alphaDW$ can reproduce disc dispersal and accretion properties.



\subsection{Relation between $\alpha$-parameters and the disc physics}
\label{subsec:numerical-simu}
Numerical simulations or semi-analytical wind solutions can be used to relate the phenomenological $\alpha$ and $\lambda$ parameters to disc physical quantities. $\alphaDW$ is certainly the most important parameter of our solutions and is roughly proportional to the magnetization of the disc, also denoted as the inverse of the $\beta_0$ parameter, where 
\begin{equation}
 \beta_0 \equiv \frac{8 \pi P_0}{B_{z,0}^2} 
 \label{eq:beta}
\end{equation}
is the thermal to the magnetic pressure ratio in the mid-plane, with $B_{z,0}$ the strength of the vertical component of the magnetic field in the mid-plane. \rev{In fact, combining Eq. (\ref{eq:Tzphi}) and (\ref{eq:alpha-DW}), $\alphaDW$ is proportional to $< B_z B_{\phi}>$ evaluated in the disc upper layers:
\begin{equation}
    \alphaDW = \frac{8}{3\sqrt{2\pi}} \frac{< B_z B_{\phi}>_{z=\pm H_W}}{4\pi\epsilon P_0},
    \label{eq:alpha-DW-BphiBz}
\end{equation}
where we assumed $< B_z B_{\phi}>_{z=+H_W}=< B_z B_{\phi}>_{z=-H_W}$.
For typical discs dominated by ambipolar or Ohmic effects, $B_z$ is roughly constant over the disc scale height whereas $B_{\phi}$ increases with altitude. The exact amount of $B_{\phi}$ generated in the atmosphere depends on the detailed microphysics of the disc. Global numerical simulations typically find $B_{\phi}(z=\pm H_W)/B_{z,0} \sim 1-20$ \citep{2002ApJ...581..988C,2017A&A...600A..75B,2017ApJ...845...75B}. Combining the latter relation and Eq. (\ref{eq:alpha-DW-BphiBz}) yields to}
\begin{equation}
    \alphaDW \simeq 2-40 \times 10^{-3} \left( \beta_0/10^4\right)^{-1} \left( \epsilon/0.1\right)^{-1}.
    \label{eq:alpha-DW-simu}
\end{equation}
The constraints on $\alphaDW$ obtained in Sec. \ref{subsec:disc-lifetime-tacc} from the observed disc lifetime translate to the typical value of the initial magnetisation of about $\beta_0 \simeq 10^5$ with a rather large spread. Using Eq. (\ref{eq:alpha-numerical-value}) in the pure wind driven case, Eq. (\ref{eq:alpha-DW-simu}) can be translated to a relation between $\tacc$ and the initial value of $\beta_0$ at $r=\rc$ :
\begin{equation}
    \tacc \simeq 0.05-1\times 10^6 yr \left( \beta_0(r_c)/10^4\right) \left(r_c/50au\right)^{5/4} \revmnras{\left(M_*/1M_{\odot}\right)^{-1/2}},
\end{equation}
where we assume $\epsilon(r) = 0.1 (r/50au)^{1/4}$ and a constant $\alpha_{DW}$ across the disc.

The values of $\alphaSS$ is highly uncertain since it depends on the ability of the disc to develop the MRI. It can thus vary by orders of magnitudes for the same disc magnetization. We also note that even in the absence of MRI, radial transport of angular momentum can occur due to various kinds of instabilities such as the vertical shear instability leading to $\alphaSS$ values of about $\sim 10^{-4}$.

\subsection{Assumption for analytical solutions}
\label{subsubsec:simplifying-assumptions}
The $\alpha$-framework defined is Sec. \ref{sec:model} provides modelers with a simple parameterization of wind accretion that can be used to compute numerical solutions, with $\alphaDW$, $\alphaSS$ and $\lambda$ parameters that varies both in space and time. In this work, in order to obtain analytical solutions, we made a number of simplifying assumptions on the radial profile and the time dependence of the phenomenological parameters. 

Following Eq. (\ref{eq:alpha-DW-simu}), the assumption of $\alphaDW$ constant across the disc amounts to assuming that the magnetisation of the disc is almost constant across the disc with $\beta_0(r) \propto \epsilon(r)^{-1} \propto r^{-1/4}$ for the assumed temperature profile of $T(r) \propto r^{-1/2}$. The extension of the solutions to $\alpha$-parameters and temperature with arbitrary power-law dependence on radius can describe any power-law distribution of $\beta_0(r)$ or $T(r)$ (see Appendix \ref{app:gamma-solutions}). 

The dependency of $\alphaDW$ with time, investigated in the pure wind case ($\alphaSS =0$), can be seen as a result of the evolution of the magnetic field strength. The secular evolution of the magnetic field in wind-driven accretion discs remains largely unknown \citep{1994MNRAS.267..235L,2014MNRAS.441..852G,2014ApJ...785..127O,2014ApJ...797..132T,2017ApJ...836...46B}. In the first class of solution, the assumption of constant $\alphaDW$ amounts to assuming that the evolution of the magnetic field is such that $\beta_0$ remains locally constant over time (see Eq. (\ref{eq:alpha-DW-simu})). The effect of the evolution of the magnetic field is highlighted by the second class of solution. For $\omega =1$, $\alphaDW \propto \Sigmac^{-1}$. Since $\alphaDW \propto B_z^2/\Sigma$, the latter relation implies that $B_z$ is locally constant over time. The solutions with $0 < \omega< 1$ describe intermediate cases for which the magnetic field strength declines slower than the surface density of the gas. Quantitatively, the magnetic flux declines with time as $\Phi_B(t) \propto \Sigmac(t)^{(1-\omega)/2}$. Overall, our finding that for $\omega \ge 0$ the disc is fully dispersed at a finite time demonstrates the decisive role of the transport of the magnetic field in the dispersal of the disc. In turn, the transport of the magnetic field can be observationally constrained from disc demographics.

\rev{Our self-similar ansatz (\ref{eq:sigma-ansatz}) assumes that the disc exhibits a power-law surface density profile tapered by an exponential cutoff. One can wonder if a different initial profile would naturally tend to these self-similar ansatz. In viscous discs, it has been shown that the  viscous spreading naturally produces the exponential cutoff assumed in the \citet{1974MNRAS.168..603L} self-similar solutions. In contrast, in pure wind-driven case, the gas is simply advected toward the star such that $\Sigma$ keeps memory of the initial conditions. The use of the ansatz (\ref{eq:sigma-ansatz}) in the wind-driven case might then appear somewhat arbitrary. One can invoke two arguments to justify the initial surface density profile: the formation of the disc can lead to this kind of profile, due to the distribution of angular momentum in the collapsing envelope, or a small amount of turbulence (non-vanishing $\alphaSS$) that would produce a smooth outer edge even if the transport of angular momentum is dominated by the wind.}

\rev{The inner radius of the disc denoted as $r_{in}$ plays an important role in our solution as it sets the wind mass loss-rate (see Eq. (\ref{eq:fM-def})).
Recalling that our disc model is designed to describe the bulk part of the disc, one has to consider $r_{in}$ as the inner radius of the wind- launching region, typically about a few $au$ and not necessarily the inner radius of the disc which is about 0.05 au in T Tauri discs. In that case, our expressions of $\dot{M}_*(t)$ are valid under the assumption that the accretion rate is constant across the innermost disc, between 0.05 and few au. In that region, the angular momentum can be either transported by turbulence or by a fast MHD disc wind observed as a jet \citep{2004A&A...416L...9P} with a low mass loss rates \citep{2018A&A...609A..87N}.}

\subsection{Comparison with previous evolution models}
\label{subsubsec:previous-models}

Over the past few years, a handful models have been proposed to describe the secular evolution of discs under the effect of an MHD disc wind \citep{2013ApJ...778L..14A,2016ApJ...821...80B,2016A&A...596A..74S,2017ApJ...845...31H,2018MNRAS.475.5059K,2019ApJ...879...98C}. As in the present work, they all rely on vertically integrated disc equations that allow us to follow the disc evolution on Myr timescales. The main difference between these models lies in the prescriptions of the wind and turbulent stress tensors, and of the local mass-loss rate (see Eq. (\ref{eq:master-eq})). The equivalence between the different parameters found in various papers are summarized in Appendix \ref{app:alphaDW-def}. 

In the present paper, we parameterize the radial and vertical stress tensors, and the mass-loss rate using the phenomenological $\alpha$ and $\lambda$ parameters rather than relying on self-consistent models of magnetized disc. Therefore, our model does not depend on the details of the physical and chemical structure of the disc nor on results of numerical simulations. This approach is very similar to that followed by \citet{2020A&A...633A...4K} in the context of planet migration but has the advantage that $\alphaDW$ has a considerably easier physical interpretation. This contrasts with the majority of the published models. In their pioneering work, \citet{2013ApJ...778L..14A}, use results of shearing-box simulations to compute the resulting torques from an assumed disc magnetization \citep[see also][]{2017ApJ...845...31H,2018MNRAS.475.5059K}. \citet{2016ApJ...821...80B} uses instead semi-analytical wind solutions that describe the launching of the wind from a warm atmosphere threatened by a magnetic field of prescribed configuration. Our approach is closer to that of \citet{2016A&A...596A..74S} who parameterize the wind torque using an $\alpha$-like parameter. In the latter study, the mass-loss rate is parameterized using sophisticated energetic arguments. The draw back of this kind of parameterization it to introduce characteristic scales related to the mass-loss rate and rely on uncertain assumptions on the wind mass-loading. Moreover, this complex parameterization prevents one from finding simple and exact analytical solutions \citep{2019ApJ...879...98C} that are key for the exploration of the parameter space and the comparison to observations.

Despite these differences, our analytical solutions capture most of the key features of the disc evolution solutions that have been qualitatively analysed using numerical solutions. In particular, \citet{2013ApJ...778L..14A} already observed the spreading of the disc in hybrid solutions. Our analytical solutions show that in the simple case of constant $\alpha$-parameters, this spreading occurs according to the viscous timescale as in the pure viscous case (see Eq. (\ref{eq:disc-radius})). \citet{2013ApJ...778L..14A}, followed by \citet{2016ApJ...821...80B} have also demonstrated that a constant magnetic flux leads to the full dispersal of the disc at finite time. Our second class of analytical solution describes a range of situation for which the magnetic flux, or equivalently, the magnetic field strength, declines at slower pace than the gas. One of the features that is not captured by our analytical solutions is the formation of a cavity predicted by \citet{2016A&A...596A..74S}. This is due to the simplifying assumption of uniform $\alpha$ and $\lambda$ parameters which does not introduce any specific scale in the solution. A more sophisticated prescription of the spatial and temporal evolution of $\alpha$ and $\lambda$, as done in \citet{2016A&A...596A..74S} \rev{can be adopted to recover these types of solutions if strong constraints are available on what the characteristic scale should be.} 


\section{Conclusions}
\label{sec:conclusions}
In this work, we present an extension of the $\alpha-$framework to describe the secular evolution of discs for which accretion is governed by an MHD disc wind. The wind torque is parameterized using an $\alphaDW$ parameter that can be readily compared to the $\alphaSS$ parameter. The mass loss rate is parameterized using the $\lambda$ parameter that can be observationally constrained. Whereas the formalism can be used to compute the evolution of a disc with $\alphaDW$, $\alphaSS$, and $\lambda$ that vary in space and time, we made a number of simplifying assumption to find analytical solutions. These analytical solutions constitute an extension of the canonical \citet{1974MNRAS.168..603L} solutions. The strength of these self-similar solutions is two fold: first, the fundamental features of wind-driven accretion are analysed in detail. In particular, we show that the absence of the disc spreading leads to a steep decline of the disc mass, is contrast with viscous accretion. If the strength of the magnetic field declines slowly with time, the evolution of the disc is even more radical as it is fully dispersed after a finite time. Secondly, these solutions open a new avenue to test MHD wind-driven accretion from the observed disc demographics. Depending on the wind torque and on the evolution of the magnetic field, a population of disc is shown to cluster is a specific way in the $\MaccMd$ plane. Interestingly, we show that wind-driven accretion can produce a broad dispersion in the $\MaccMd$ plane and a correlation between $\MaccMd$ depending on the initial condition of the disc population. Observationally constrained disc evolution models will be crucial to build realistic planet formation models in the emerging paradigm of MHD disc winds.

\section*{Acknowledgements}

\revmnras{We thank the anonymous referee for their careful reading of the manuscript}. B.T. acknowledges support from the research programme Dutch Astrochemistry Network II with project number 614.001.751, which is (partly) financed by the Dutch Research Council (NWO). G.R. acknowledges support from the Netherlands Organisation for Scientific Research (NWO, program number 016.Veni.192.233) and from an STFC Ernest Rutherford Fellowship (grant number ST/T003855/1).

\section*{Data Availability}

No new data were generated or analysed in support of this research.




\bibliographystyle{mnras}
\bibliography{export-bibtex} 




\appendix
\section{Master equation}
\label{app:eq-basics}
In this section, the master equation (\ref{eq:master-eq-final}) is derived from the basic MHD equations.
The conservation of angular momentum in cylindrical coordinates writes
\begin{equation}
\begin{split}
     \partial_t(\rho r \varv_{\phi}) = & -\frac{1}{r} \partial_r \left\{ r^2 \left( \rho \varv_r \varv_{\phi} -B_rB_{\phi}/4\pi \right)\right\} \\
    & -\partial_z \left\{ r \left( \rho \rev{\varv_{\phi}} \varv_z -B_z B_{\phi}/4\pi \right)\right\}. 
    \label{eq:cons-AM-2D}
\end{split}
\end{equation}
We further assume that the disc is geometrically thin, isothermal, and in nearly Keplerian rotation, decomposing the azimutal velocity as $\varv_{\phi} = r \Omega + \delta \varv_{\phi}$. Equation (\ref{eq:cons-AM-2D}), integrated between the top and the bottom surface of the disc, yields
\begin{equation}
\begin{split}
     \partial_t(\Sigma r^2 \Omega) = & -\frac{1}{r} \partial_r \left\{ r^3 \Omega \Sigma \varv_r \right\}  -\frac{1}{r} \partial_r \left\{ r^2 \int^{+H_W}_{-H_W} T_{r\phi} dz \right\} \\
    & -r^2 \Omega \dot{\Sigma}_W -r | T_{z\phi}|^{+H_W}_{-H_W},
    \label{eq:cons-AM-1D}
\end{split}
\end{equation}
\rev{where $\dot{\Sigma}_W$ is the wind mass loss rate per unit surface,} and $T_{r\phi}$ and $T_{z\phi}$ are the components of the stress tensor associated to the radial and vertical transport of angular momentum. Their expressions are given in Eq. (\ref{eq:Trphi}) and (\ref{eq:Tzphi}). $H_W$ denotes the typical scale height from which the wind is launched, which is about a few times the hydrostatic scale height.

The conservation of the mass writes
\begin{equation}
\begin{split}
     \partial_t \rho = & -\frac{1}{r} \partial_r \left[ r \rho \varv_r\right] -\partial_z \left[ \rho \varv_z \right],
    \label{eq:cons-mass-2D}
\end{split}
\end{equation}
which leads to the vertically integrated form
\begin{equation}
\begin{split}
     \partial_t \Sigma = & -\frac{1}{r} \partial_r \left[ r \Sigma \varv_r\right] -\dot{\Sigma}_W.
    \label{eq:cons-mass-1D}
\end{split}
\end{equation}
The advection velocity $\varv_r$ is set by the extraction of angular momentum. Combining Eq. (\ref{eq:cons-AM-1D}) and (\ref{eq:cons-mass-1D}), we obtain the master equation
\begin{equation}
\begin{split}
       \partial_t\Sigma = \frac{2}{r}\partial_r \left\{ \frac{1}{r\Omega} \partial_r \left(r^2  \int_{-H_W}^{+H_W} T_{r\phi} dz \right)\right\}
       + \frac{2}{r} \partial_r\left\{\frac{r | T_{z\phi}|^{+H_W}_{-H_W}}{\Omega} \right\}
       - \dot{\Sigma}_W ,
\end{split}
\label{eq:master-eq-app}
\end{equation}
which describes the evolution of the surface density under the action of the radial and vertical toques, and the wind mass loss. In this work, the torques are parameterized using the dimensionless parameters $\alphaSS$ and $\alphaDW$ that can be straightforwardly injected in Eq. (\ref{eq:master-eq-app}).

In order to relate the wind mass loss $\dot{\Sigma}_W$ to the wind torque, we further assume that the angular momentum of the wind per unit mass is proportional to that in the disc as quantified by the magnetic lever arm parameter $\lambda$. The total angular momentum lost locally by the disc per unit surface is then $-\lambda r^2 \Omega \dot{\Sigma}_W$, and is equal to the last two terms of Eq. (\ref{eq:cons-AM-1D}). Therefore, the local mass loss rate is
\begin{equation}
\begin{split}
     \dot{\Sigma}_W &= \frac{r| T_{z\phi}|^{+H_W}_{-H_W}}{\Omega r^2 (\lambda-1)} = \frac{3 \alphaDW \Sigma c_s^2}{4 (\lambda-1)\Omega r^2},
    \label{eq:sigmaW-cons-AM}
\end{split}
\end{equation}
where we adopt the definition of $\alphaDW$ provided in Eq. (\ref{eq:alpha-DW}). Injecting this relation in Eq. (\ref{eq:master-eq-app}), we obtain the master equation (\ref{eq:master-eq-final}).

\section{Equivalence between published disc models}
\label{app:alphaDW-def}

Over the past few years, a handful of disc evolution models have been constructed, with various assumptions and using different dimensionless parameters to describe the wind torque and the mass-loss rate. In this subsection, we provide the relation between these parameters and our $\alphaDW$ and $\lambda$ parameters to facilitate the comparison with previously published models.

The wind torque is at the base of any disc evolution model. The normalized accretion stress $W_{r\phi}$ is often defined as \citep[e.g.][]{2013ApJ...778L..14A,2017ApJ...845...31H}
\begin{equation}
  W_{z\phi}\equiv \frac{| T_{z\phi}|^{+H_W}_{-H_W}}{ 2\rho_0 c_s^2}  = \frac{3\sqrt{2\pi}}{8} \epsilon \alphaSS \simeq \epsilon \alphaDW.
\end{equation}
or by others \citep{2016A&A...596A..74S} as
\begin{equation}
  \bar{\alpha}_{z\phi} \equiv \rev{2 W_{z\phi}} = \frac{3\sqrt{2\pi}}{4} \epsilon \alphaDW.
  \label{eq:app-suzuki-alphaDW}
\end{equation}
\citet{2021A&A...650A..35L} uses the parameter
\begin{equation}
    \upsilon_{\pm} \equiv \pm \frac{T_{z \phi}(z=\pm H_w)}{\Sigma \Omega^2 H} = \frac{3}{8} \epsilon \alphaDW.
\end{equation}
In their study about planet migration, \citet{2020A&A...633A...4K} do not parameterize the wind torque directly but the local mass loss rate with a parameter $b$, and assume a constant magnetic lever arm parameter $\lambda$. This leads to the conversion
\begin{equation}
  b \equiv \frac{2 \pi \dot{\Sigma}_W}{\Omega \Sigma} =  \frac{3\pi}{2 (\lambda-1)} \epsilon^2 \alphaDW.
\end{equation}

The effect of the wind mass loss rate in often neglected \citep[e.g.][]{2013ApJ...778L..14A,2017ApJ...845...31H}. \citet{2016A&A...596A..74S} adopt the parameter
\begin{equation}
  C_W \equiv \frac{\dot{\Sigma}_W}{\rho_0 c_s} = \frac{3\sqrt{2\pi}}{4(\lambda-1)} \epsilon^2 \alphaDW,
  \label{eq:app-suzuki-Cw}
\end{equation}
which is computed from considerations about the energetics of the flow. \citet{2021A&A...650A..35L} uses instead the parameter
\begin{equation}
  \zeta_{\pm} \equiv \frac{\dot{\Sigma}_W}{2 \Sigma \Omega} = \frac{3}{8(\lambda-1)} \epsilon^2 \alphaDW = \frac{\epsilon \upsilon_{\pm}}{(\lambda-1)},
\end{equation}
assuming that the enthalpy of the wind is negligible.

\citet{2019ApJ...879...98C} uses 3 parameters to describe the impact of turbulence and the disc wind on the evolution of the disc. The equivalence between our parameters and theirs can be found using the parameters of \citet{2016A&A...596A..74S} (see Eq. (\ref{eq:app-suzuki-alphaDW}) and (\ref{eq:app-suzuki-Cw})) and their Eq. (5).

\section{Extension of the hybrid solutions to other power-laws}
\label{app:gamma-solutions}
Assuming that $\alphaSS$ and $\alphaDW$ are constant in time and scale with radius as $\alphaSS  c_s^2 \propto r^{-3/2+\gamma}$ and $\alphaDW c_s^2 \propto r^{-3/2+\gamma}$, the master equation Eq. (\ref{eq:master-eq-final}) can written using the dimensionless coordinates $\tilde{r} = r/r_{c}(0)$ and $\tilde{t}= t/t_{\nu,0}$ as
\begin{equation}
\begin{split}
(2-\gamma)^2 \partial_{\tilde{t}} \Sigma(\tilde{r},\tilde{t}) & = 
~\tilde{r}^{-1} \partial_{\tilde{r}} \left( \tilde{r}^{1/2} \partial_{\tilde{r}} (\tilde{r}^{1/2+\gamma} \Sigma(\tilde{r},\tilde{t})) \right) \\
+ & \frac{\psi}{2} \tilde{r}^{-1} \partial_{\tilde{r}} \{\tilde{r}^{\gamma} \Sigma(\tilde{r},\tilde{t}) \} -\frac{\psi}{4 (\lambda-1)} \tilde{r}^{-2+\gamma} \Sigma(\tilde{r},\tilde{t}),
\end{split}
    \label{eq:master-eq-appendix-red}
\end{equation}
where the viscous timescale is
\begin{equation}
    t_{\nu,0} = \frac{r_c}{3 (2-\gamma)^2 \epsilon_c c_{s,c} \alpha_{SS,c}}.
\end{equation}
and $r_{\rm{c}}$ is an arbitrary radius. We also define the initial accretion timescale as 
\begin{equation}
    \tacc = \frac{r_c}{3 (2-\gamma)^2 \epsilon_c c_{s,c} \tilde{\alpha}_{c}}.
\end{equation}

The steady-state solution is found by assuming that the surface density follows a power-law dependence with radius. This leads to $\Sigma \propto r^{\xi-\gamma}$, where $\xi$ is the mass ejection index given in Eq. (\ref{eq:xi}).

Inspired by the steady-state solution and by the self-similar solution of \citet{1974MNRAS.168..603L} for an arbitrary value of $\gamma$, we find exact solutions of Eq. (\ref{eq:master-eq-appendix-red}) using the ansatz
$\Sigma(\tilde{r},\tilde{t}) = A(\tilde{t}) \tilde{r}^{-\gamma+\xi} e^{-(\tilde{r}/\tilde{r}_{\rm{c}}(\tilde{t}))^{2-\gamma}},$
where $\tilde{r}_{\rm{c}}(\tilde{t})$ is the characteristic disc radius, $\xi$ is the mass ejection index defined in Eq. (\ref{eq:xi}), and $A(\tilde{t})$ is a function of time only. Injecting this form in Eq. (\ref{eq:master-eq-appendix-red}) leads to a system of two equations
\begin{equation}
\begin{split}
\dot{r}_c(t) & = \frac{1}{2-\gamma}r_c(t)^{\gamma-1}, \\
\dot{A}(t) & = \frac{1}{2-\gamma} \left( \gamma -\frac{5}{2} -2\xi - \frac{\psi}{2} \right) r_c(t)^{\gamma-2} A(t).
\end{split}
\end{equation}
In dimensional form, the solution of the system is
\begin{equation}
\begin{split}
r_c(t) & = r_c(0) (1+t/t_{\nu, 0})^{1/(2-\gamma)}, \\
A(t) & = A(0) \left(1+t/t_{\nu, 0} \right)^{-(5/2-\gamma+2\xi+\psi/2)/(2-\gamma)}.
\end{split}
\end{equation}
Rewritting the ansatz as
\begin{equation}
\Sigma(r,t) = \Sigmac(t) (r/r_c(t))^{-\gamma+\xi} e^{-(r/r_{\rm{c}}(t))^{2-\gamma}},
\end{equation}
we find the solutions
\begin{equation}
\begin{split}
r_c(t) & = r_c(0) \left(1+\frac{t}{(1+\psi)\tacc}\right)^{1/(2-\gamma)}, \\
\Sigmac(t) & = \Sigmac(0) \left(1+\frac{t}{(1+\psi)\tacc}\right)^{-(5+2\xi+\psi)/(2(2-\gamma))}.
\end{split}
\end{equation}
The disc mass is then obtained using Eq. (\ref{eq:app-disc-mass-gamma})
\begin{equation}
    M_D(t) = M_0 \left(1+\frac{t}{(1+\psi)\tacc}\right)^{-(1+2\xi+\psi)/2(2-\gamma)}.
    \label{eq:app-MD-gamma}
\end{equation}
Following the case of $\gamma =1$ detailed in Appendix \ref{app:fM}, one can also show that the mass ejection-to-accretion ratio $f_M(t)$ does not depends on $\gamma$. Therefore the accretion rate can be derived from Eqs. (\ref{eq:dotM*-def}), (\ref{eq:fM-def}), and (\ref{eq:app-MD-gamma}), which gives
\begin{equation}
\begin{split}
        \dot{M}_*(t) & =  \dot{M}_{*,0}  \left(1+\frac{t}{(1+\psi)\tacc}\right)^{-(5-2\gamma+4\xi+\psi)/2(2-\gamma)}, \\
        \dot{M}_{*,0} & = \frac{1+2\xi +\psi}{\psi+1} \frac{M_0}{2(2-\gamma)\tacc (1+f_{M,0})}.
\end{split}
\end{equation}

\section{Disc mass}
\label{app:disc-mass}
The disc mass for both classes of solution can be obtained by integrating the surface density over the full extent of the disc. For $\alpha$ parameters constant in space, this gives
\begin{equation}
\begin{split}
     M_D(t)& = 2\pi \Sigmac(t) r_c(t)^2 \int_{r_{in}/r_c(t)}^{+\infty} x^{\xi} e^{-x} dx \\
     & \simeq 2\pi r_c(t)^2 \Sigmac(t) \Gamma(\xi+1),
     \end{split}
\end{equation}
where $\Gamma$ is the gamma function and where we assume $r_{in} \ll r_c(t)$. Interestingly, for $\lambda > 3/2$, $\xi$ \revmnras{is smaller} than unity and $\Gamma(\xi+1)$ ranges from 0.88 and 1. In this work, we assume $\Gamma(\xi+1) \simeq 1$.

For $\alpha$-parameters and a temperature that scales as $\alpha c_s^2 \propto r^{-3/2+\gamma}$,
\begin{equation}
    M_D(t)  =  2\pi \Sigmac(t) r_c(t)^2 \int_{r_{in}/r_c(t)}^{+\infty} x^{1-\gamma+\xi} e^{-x^{2-\gamma}} dx.
\end{equation}
By change of variable and assuming $r_c \gg r_{in}$, we obtain
\begin{equation}
    \begin{split}
    M_D(t)  \simeq & \frac{2\pi \Sigmac(t) r_c(t)^2}{2-\gamma} \int_{0}^{+\infty} x^{\xi/(2-\gamma)} e^{-x} dx \\
    \simeq & \frac{2\pi \Sigmac(t) r_c(t)^2}{2-\gamma} \Gamma\left(\frac{\xi+2-\gamma}{2-\gamma}\right).
    \label{eq:app-disc-mass-gamma}
     \end{split}
\end{equation}

\section{Mass ejection-to-accretion ratio}
\label{app:fM}
In Sec. \ref{sec:anal-sol}, we defined for convenience the stellar accretion rate and the mass loss-rate using the mass ejection-to-accretion ratio $f_M(t) = \dot{M}_W(t)/\dot{M}_*(t)$. Here, we show that for the two classes of analytical solutions presented in this work, $f_M(t) = (r_c(t)/r_{in})^{\xi}-1$.

The total stellar accretion rate is the sum of the accretion rate due to viscosity and due to the wind evaluated at $r=\rin$:
\begin{equation}
    \dot{M}_* = \dot{M}_{acc}^{visc}(\rin) + \dot{M}_{acc}^{DW}(\rin).
\end{equation}
At any time, the surface density profile is $\Sigma(r,t) = \Sigmac(t) (r/\rc(t))^{-1+\xi} e^{-r/\rc(t)}$ and Eqs. (\ref{eq:Maccvisc}) and (\ref{eq:MaccW}) yield 
\begin{equation}
\begin{split}
   \dot{M}_{acc}^{W}(\rin) & = \frac{\psi}{1+\psi} \frac{M_D(t)}{2 \tacct} (\rin/\rc(t))^{\xi}, \\
   \dot{M}_{acc}^{visc}(\rin) & = \frac{2\xi+1}{1+\psi} \frac{M_D(t)}{2\tacct} (\rin/\rc(t))^{\xi},
\end{split}
\end{equation}
where $\tacct$ is the instantaneous accretion timescale defined in Eq. (\ref{eq:tacc-instant-def}) and where we assume $\rin \ll \rc(t)$.
The stellar accretion rate is then
\begin{equation}
\dot{M}_*(t) = \frac{\psi + 2\xi +1}{\psi+1} \frac{M_D(t)}{2\tacct} (\rin/\rc(t))^{\xi}.
\label{eq:app-M*}
\end{equation}

With the self-similar ansatz, the local wind mass loss rate given by Eq. (\ref{eq:sigmaW-cons-AM}) is
\begin{equation}
    \dot{\Sigma}_W(r,t) = \frac{\psi}{2(\lambda-1)(\psi+1)} \frac{\Sigmac(t)}{2\tacct} \left( \frac{r}{\rc(t)} \right)^{-2+\xi} e^{-r/\rc(t)}.
\end{equation}
and the total mass-loss rate is
\begin{equation}
       \dot{M}_W = 2\pi \int_{\rin}^{+\infty} \dot{\Sigma}(r,t) r dr.
    \label{eq:app-def-dotMW}
\end{equation}
Combining the latter two equations the total mass loss rate writes
\begin{equation}
\begin{split}
    \dot{M}_W = \frac{\psi}{2(\lambda-1)(\psi\revmnras{+1})} \frac{M_D(t)}{2 \tacct} \int_{0}^{+\infty} x^{-1+\xi} e^{-x} dx,
     \end{split}
    \label{eq:app-def-dotMW-2}
\end{equation}
where we assumed $\rin \ll r_c(t)$ and use the relation $M_D(t)=2\pi \Sigmac(t) r_c(t)^2$. The integral $G(\xi)= \int_{\rin/\rc}^{+\infty} x^{-1+\xi} e^{-x} dx$ can be computed using integration by part of the type $\int u'(x)v(x)dx=-\int u(x)v'(x)dx+[u(x)v(x)]$ with
\begin{equation}
\begin{split}
    u(x)&=x~~~~~~~~~~~~~~u'(x) =1\\
    v(x)&=x^{-1+\xi}e^{-x} ~~~ v'(x) =-v(x)+(-1+\xi)x^{-2+\xi}e^{-x},
\end{split}
\end{equation}
This gives
\begin{equation}
  G(\xi) = \int_{\rin/\rc}^{+\infty} x^{\xi} e^{-x} - (-1+\xi) G(\xi) - (\rin/\rc)^{\xi}.
\end{equation}
Adopting $\rin/\rc \ll 1$ in the first RHS term and rearanging the terms gives
\begin{equation}
  G(\xi) = \frac{1}{\xi} (\Gamma(\xi+1) - (\rin/\rc)^{\xi}).
\end{equation}
As for the disc mass, we adopt $\Gamma(\xi+1)\simeq 1$. The total mass loss rate (\ref{eq:app-def-dotMW-2}) can then we re-written as
\begin{equation}
\begin{split}
    \dot{M}_W = \frac{\psi}{4(\lambda-1)(\psi+1)\xi \tacct} M_D(t) (1-(\rin/\rc(t))^{\xi}).
     \end{split}
    \label{eq:app-def-dotMW-3}
\end{equation}

Combining Eqs. (\ref{eq:app-M*}) and (\ref{eq:app-def-dotMW-3}) yields to
\begin{equation}
f_M(t) = \frac{\psi}{2 \xi (\lambda-1) (1+2\xi+\psi)} \left( \left( \frac{\rc(t)}{\rin} \right)^{\xi} -1 \right).
    \label{eq:app-fM-1}
\end{equation}
From the quadratic equation (\ref{eq:xi-equation}) that sets the expression of $\xi$, one can note that $(1+2\xi +\psi)\xi  = \psi / \left[ 2(\lambda-1) \right]$. This simplifies greatly Eq. (\ref{eq:app-fM-1}) and yields to 
\begin{equation}
f_M(t) = \left( \frac{\rc(t)}{\rin} \right)^{\xi} -1.
    \label{eq:app-fM-2}
\end{equation}

\bsp	
\label{lastpage}
\end{document}